\documentstyle[12pt]{article}

\def\ar{\rightarrow}
\def\bib{\bibitem}
\def\dim{\,\mbox{dim}\,}
\def\Det{\,\mbox{Det}\,}

\def\Dsl{D\!\!\!\!/}
\def\intx{\int\! d^{\sl 4}x}
\def\intX{\int\! d^{D}X\,}
\def\intp{\int\! \frac{d^{\sl 4}p}{(2{\pi})^{\sl 4}}}
\def\intP{\int\! \frac{d^{D}P}{(2{\pi})^D}\,}

\def\lar{\longrightarrow}
\def\Ln{\mbox{Ln}}
\def\pa{\partial}
\def\rvec{\!\!\!\!^{^\rightarrow}}
\def\lvec{\!\!\!\!^{^\leftarrow}}
\def\TA{\:\tilde{\!\!A}}

\def\Tr{\,\mbox{Tr}\,}

\def\al{\alpha}
\def\be{\beta}
\def\ga{\gamma}
\def\de{\delta}
\def\ep{\varepsilon}
\def\ze{\zeta}

\def\la{\lambda}
\def\va{\varphi}
\def\si{\sigma}
\def\om{\omega}

\def\Ga{{\it\Gamma}}
\def\La{{\it\Lambda}}
\def\Om{{\it\Omega}}

\def\Pit{{\it\Pi}}
\def\TPit{\:\tilde{\!\!\Pit}}
\def\Si{\Sigma}

\def\beq{\begin{equation}}
\def\eeq{\end{equation}}
\def\bed{\begin{displaymath}}
\def\eed{\end{displaymath}}
\def\beqq{\begin{eqnarray}}
\def\eeqq{\end{eqnarray}}
\def\bedd{\begin{eqnarray*}}
\def\eedd{\end{eqnarray*}}

\begin{document}

\centerline{\normalsize\bf QUANTUM ISOMETRODYNAMICS}

\vspace*{0.9cm}
\centerline{\footnotesize C. WIESENDANGER}
\baselineskip=12pt
\centerline{\footnotesize\it Aurorastr. 24, CH-8032 Zurich}
\centerline{\footnotesize E-mail: christian.wiesendanger@zuerimail.com}

\vspace*{0.9cm}
\baselineskip=13pt
\abstract{Classical Isometrodynamics is quantized in the Euclidean plus axial gauge. The quantization is then generalized to a broad class of gauges and the generating functional for the Green functions of Quantum Isometrodynamics (QID) is derived. Feynman rules in covariant Euclidean gauges are determined and QID is shown to be renormalizable by power counting. Asymptotic states are discussed and new quantum numbers related to the "inner" degrees of freedom introduced. The one-loop effective action in a Euclidean background gauge is formally calculated and shown to be finite and gauge-invariant after renormalization and a consistent definition of the arising "inner" space momentum integrals. Pure QID is shown to be asymptotically free for all dimensions of "inner" space $D$ whereas QID coupled to the Standard Model fields is not asymptotically free for $D\leq 7$. Finally nilpotent BRST transformations for Isometrodynamics are derived along with the BRST symmetry of the theory and a scetch of the general proof of renormalizability for QID is given.}

\normalsize\baselineskip=15pt

\section{Introduction}
In the search of a consistent new type of perturbatively renormalizable and unitary gauge field theory in four spacetime dimensions we have developed Isometrodynamics in \cite{chw2}. This is the gauge theory of the group ${\overline{DIFF}}\,{\bf R}^D$ of volume-preserving diffeomorphisms of a $D$-dimensional "inner" space $({\bf R\/}^{D},g)$ endowed with a flat metric $g$.

We have formulated Classical Isometrodynamics in both the Lagrangian and the Hamiltonian frameworks and demonstrated that the theory can be set up with a rigour similar to that achieved for classical Yang-Mills gauge theories even though the gauge group in our case is not compact which brings along additional challenges.

In the following we develop Quantum Isometrodynamics and show that it can be formulated as a renormalizable and asymptotically free gauge field theory in analogy to Yang-Mills theories of compact Lie groups \cite{stw2,stp,cli,tpc}.

The notations and conventions used are given in Appendix C.

\section{Quantization in the Euclidean plus Axial Gauge}
In this section we quantize Isometrodynamics in the Euclidean plus axial gauge deriving path integrals for Green functions with a manifestly invariant gauge field weight and an invariant functional measure restricted to the relevant gauge algebra ${\overline{\bf diff}}\,{\bf R}^D$.

Let us start with Hamiltonian Isometrodynamics in the Euclidean plus axial gauge $A_{\sl 3}\,\!^M =0$ with Cartesian coordinates and Euclidean metric $\de$ in "inner" space as developed in \cite{chw2}.

The $D\times 2$ independent canonical variables of the theory in this gauge are $A_i\,^M$ and their conjugates $\Pit_j\,\!^N$ for $i,j = {\sl 1,2}$. They are subject to the constraints
\beqq \label{1}
\nabla_M A_i\,^M &=& 0 \nonumber \\
\nabla^N \Pit_{j\,N} &=& 0 
\eeqq
which assure that the corresponding $A_i = A_i\,^M \nabla_M$ and $\Pit_j = \Pit_j\,\!^N \nabla_N$ are elements of the gauge algebra ${\overline{\bf diff}}\,{\bf R}^D$ of infinitesimal volume-preserving coordinate transformations of ${\bf R}^D$.

We next define an expression $A_{\sl 0}\,^N$ non-local in the independent variables $A_i\,^M$, $\Pit_j\,\!^N$ - at this point just to keep the formulae below simple 
\beq \label{2}
A_{\sl 0}\,^M \equiv \frac{1}{\pa_{\sl 3}\,\!^2} \frac{1}{\La } 
\sum_{i={\sl 1}}^{\sl 2} {\cal D}_i^M\,\!_N \Pit_i\,\!^N, \eeq
where $\La$ and the covariant derivative
\beq \label{3}
{\cal D}_i^M\,\!_N = \pa_i \,\de^M\,\!_N + A_i\,^K\cdot \nabla_K \,\de^M\,\!_N
- \nabla_N A_i\,^M \eeq
have been introduced in \cite{chw2}. $A_{\sl 0}\,^N$ fullfills
\beq \label{4}
\nabla_M A_{\sl 0}\,^M = 0 \eeq
which is easily proven using the Eqn.(\ref{1}).

The Hamiltonian density ${\cal H}_{ID}$ of the theory is given by \cite{chw2}
\beqq \label{5} {\cal H}_{ID}
&\equiv& \La\, \sum_{i={\sl 1}}^{\sl 2} \Pit_{iM}\cdot {\cal D}_i^M\,\!_N
A_{\sl 0}\,^N + \frac{1}{2}\, \sum_{i={\sl 1}}^{\sl 2} \Pit_i\,\!^M \cdot \Pit_{i\,M} \nonumber \\
&+&  \frac{\La^2}{4}\, \sum_{i,j={\sl 1}}^{\sl 2} F_{ij}\,^M \cdot F_{ij\,M} \\
&+& \frac{\La^2}{2}\, \sum_{i={\sl 1}}^{\sl 2} \pa_{\sl 3}A_i\,^M\cdot 
\pa_{\sl 3}A_{i\,M} - \frac{\La^2}{2}\, \pa_{\sl 3}A_{\sl 0}\,^M\cdot 
\pa_{\sl 3}A_{{\sl 0}\,M} \nonumber \eeqq
in terms of the $A_{\sl 0}\,^M$ defined above and the expressions
\beq \label{6}
F_{ij}\,^M = \pa_i A_j\,^M - \pa_j A_i\,^M + A_i\,^N \cdot \nabla_N A_j\,^M - A_j\,^N \cdot \nabla_N A_i\,^M.
\eeq

${\cal H}_{ID}$ together with the Hamiltonian density ${\cal H}_M = \sum_n \pi_n\cdot \pa_{\sl 0} \psi_n - {\cal L}_M$ for generic "matter" fields $\psi_m$ with conjugates $\pi_n$ is our starting point for path integral quantization.

Green functions are defined as path integrals over $A_i\,^M$, $\Pit_j\,\!^N$, $\psi_m$, $\pi_n$ with gauge and "matter" field measures 
\beqq \label{7}
& & \Pi_{\!\!\!\!\!\!_{_{_{x,X,m}}}}\!\!\!\!d\psi_m \cdot
\Pi_{\!\!\!\!\!\!_{_{_{x,X,M,i={\sl 1},{\sl 2}}}}} \!\!\!\!\!\!\!\!\!\!\!\!\!\!dA_i\,^M \;
\Pi_{\!\!\!\!\!_{_{_{i={\sl 1},{\sl 2}}}}} \!\!\de
(\nabla_M A_i\,^M) \nonumber \\
& & \cdot \Pi_{\!\!\!\!\!\!_{_{_{x,X,n}}}}\!\!\!\!d\pi_n \cdot \Pi_{\!\!\!\!\!\!_{_{_{x,X,N,j={\sl 1},{\sl 2}}}}}
\!\!\!\!\!\!\!\!\!\!\!\!\!\!d\Pit_j\,^N \;
\Pi_{\!\!\!\!\!_{_{_{j={\sl 1},{\sl 2}}}}} \!\!\de (\nabla^N \Pit_{j\,N})
\eeqq
and weight
\beq \label{8}
\exp\,i\,\int\left\{\La \, \sum_{i={\sl 1}}^{\sl 2} \Pit_{i\,M}\cdot
\pa_{\sl 0} A_i\,^M - {\cal H}_{ID}+ \sum_n \pi_n\cdot \pa_{\sl 0} \psi_n - {\cal H}_M \right\}.
\eeq 
Note that the $\de$-functions in the integration measures above ensure that we integrate over gauge fields and their conjugates belonging to the gauge algebra ${\overline{\bf diff}}\,{\bf R}^D$ only.

To turn these path integrals into Lorentz-invariant expressions we apply the usual trick treating $A_{\sl 0}\,^M$ as a new independent variable which we can integrate over. The trick still works with the restricted measure Eqn.(\ref{7}). In fact, as the weight factor Eqn.(\ref{8}) is at most quadratic in $A_{\sl 0}\,^M$ we find that
\beqq \label{9}
& & \quad \int \Pi_{\!\!\!\!\!\!_{_{_{x,X,M}}}}
\!\!\!\!dA_{\sl 0}\,^M \;
\, \de (\nabla_M A_{\sl 0}\,^M) \nonumber \\
& & \cdot \exp\,i\,\int\left\{\La \, \sum_{i={\sl 1}}^{\sl 2} \Pit_{iM}\cdot
\pa_{\sl 0} A_i\,^M - {\cal H}_{ID} \right\} \nonumber \\
& & \propto \int \Pi_{\!\!\!\!\!\!_{_{_{x,X,M}}}}
\!\!\!\!d\TA_{\sl 0}\,^M \;
\, \de (\nabla_M \TA_{\sl 0}\,^M) \\
& & \cdot \exp -\frac{i}{2}\,\int \, \La^2 \, \TA_{\sl 0}\,^N \cdot
\pa_{\sl 3}\,\!\!^2 \TA_{\sl 0\,N} \nonumber \\
& & \cdot \exp \,i\,\int \frac{1}{2}\, \sum_{i,j={\sl 1}}^{\sl 2} 
{\cal D}_i^M\,_K \Pit_i\,\!^K \cdot \frac{1}{\pa_{\sl 3}\,\!^2}\,
{\cal D}_{jM}\,^L \Pit_{jL} +\dots \nonumber \\
& & \propto \exp\,i\,\int \left\{ \La \, \sum_{i={\sl 1}}^{\sl 2} \Pit_{iM}\cdot
\pa_{\sl 0} A_i\,^M - {\cal H}_{ID} \right\}
\nonumber \eeqq
after a shift of integration variables $\TA_{\sl 0}\,^M \equiv A_{\sl 0}\,^M - \frac{1}{\pa_{\sl 3}\,\!^2} \frac{1}{\La } \sum_{i={\sl 1}}^{\sl 2} {\cal D}_i^M\,\!_K \Pit_i\,\!^K$. This is - apart from a field-independent normalization factor - the gauge weight factor Eqn.(\ref{8}) with $A_{\sl 0}\,^M$ given by Eqn.(\ref{2}) in terms of $A_i\,^M$, $\Pit_j\,\!^N$. 

Next, as ${\cal H}_{ID}$ is quadratic in $\Pit_j\,\!^N$ we can perform the corresponding $\Pit_j\,\!^N$ integrations for fixed $A_i\,^M$ and $A_{\sl 0}\,^M$ and find after a shift of integration variables $\:\TPit_j\,^M \equiv \Pit_j\,^M - \La\, F_{{\sl 0}j}\,^M$
\beqq \label{10}
& & \quad \int \Pi_{\!\!\!\!\!\!_{_{_{x,X,M,j={\sl 1},{\sl 2}}}}}
\!\!\!\!\!\!\!\!\!\!\!\!\!\!d\Pit_j\,^M \;
\Pi_{\!\!\!\!\!_{_{_{j={\sl 1},{\sl 2}}}}} \!\!\de (\nabla^M \Pit_{j\,M}) \nonumber \\
& & \cdot \exp\,i\,\int \left\{-\frac{1}{2}\, \sum_{i={\sl 1}}^{\sl 2} \Pit_i\,\!^M \cdot \Pit_{i\,M} + \La\,\sum_{i={\sl 1}}^{\sl 2} F_{{\sl 0}i}\,^M \cdot \Pit_{i\,M} 
+\dots \right\} \nonumber \\
& & \propto \int \Pi_{\!\!\!\!\!\!_{_{_{x,X,M,j={\sl 1},{\sl 2}}}}}
\!\!\!\!\!\!\!\!\!\!\!\!\!\!d\TPit_j\,^M \;
\Pi_{\!\!\!\!\!_{_{_{j={\sl 1},{\sl 2}}}}} \!\!\de (\nabla^M \TPit_{j\,M}) \\
& & \cdot \exp\,i\,\int \left\{-\frac{1}{2}\, \sum_{i={\sl 1}}^{\sl 2} \TPit_i\,\!^M \cdot \TPit_{i\,M} + \int \frac{\La^2 }{2}\, \sum_{i={\sl 1}}^{\sl 2} F_{{\sl 0}i}\,^M \cdot F_{{\sl 0}i\,M} +\dots\right\}  \nonumber \\
& & \propto \exp \,i\,\int \frac{\La^2 }{2}\, \sum_{i={\sl 1}}^{\sl 2}
F_{{\sl 0}i}\,^M \cdot F_{{\sl 0}i\,M} +\dots\:.
\nonumber \eeqq
Above we have introduced
\beq \label{11}
F_{{\sl 0}i}\,^M = \pa_{\sl 0} A_i \,^M - \pa_i A_{\sl 0}\,^M + A_{\sl 0}\,^N \cdot \nabla_N A_i\,^M - A_i\,^N \cdot \nabla_N A_{\sl 0}\,^M 
\eeq
and used that $ F_{{\sl 0}i}\,^M$ is an element of the gauge algebra ${\overline{\bf diff}}\,{\bf R}^D$ as is easily verified.

As a result Green functions are given as path integrals over $A_i\,^M$, $A_{\sl 0}\,^N$, $\psi_m$ - assuming that the integrations over $\pi_n$ are Gaussian as well - with the gauge field measure 
\beq \label{12}
\Pi_{\!\!\!\!\!\!_{_{_{x,X,M}}}}
\!\!\!\!dA_{\sl 0}\,^M \;
\, \de (\nabla_M A_{\sl 0}\,^M)
\cdot \Pi_{\!\!\!\!\!\!_{_{_{x,X,M,i={\sl 1},{\sl 2}}}}} \!\!\!\!\!\!\!\!\!\!\!\!\!\!dA_i\,^M \;
\Pi_{\!\!\!\!\!\!_{_{_{i={\sl 1},{\sl 2}}}}} \!\!\de (\nabla_M A_i\,^M)
\eeq
and gauge field weight
\beqq \label{13}
& &\!\!\!\!\!\!\!\!\exp\,i\,\int\Bigg\{\frac{\La^2}{2}\,\sum_{i={\sl 1}}^{\sl 2}
F_{{\sl 0}i}\,^M \cdot F_{{\sl 0}i\,M}
- \frac{\La^2}{4}\, \sum_{i,j={\sl 1}}^{\sl 2} F_{ij}\,^M \cdot F_{ij\,M} \nonumber \\
& & -\frac{\La^2}{2}\, \sum_{i={\sl 1}}^{\sl 2} \pa_{\sl 3}A_i\,^M\cdot 
\pa_{\sl 3}A_{i\,M} + \frac{\La^2}{2}\, \pa_{\sl 3}A_{\sl 0}\,^M\cdot 
\pa_{\sl 3}A_{{\sl 0}\,M} \Bigg\}.
\eeqq

Introducing the variable $A_{\sl 3}\,^M$ which vanishes identically in the Euclidean plus axial gauge the integrals can finally be recast in a manifestly Lorentz-invariant fashion with gauge field measure
\beq \label{14}
\Pi_{\!\!\!\!\!\!_{_{_{x,X,M,\mu}}}}
\!\!\!\!\!\!\!\!dA_\mu\,^M \;
\Pi_{\!\!\!\!\!_{_{_{\mu}}}} \,\, \de (\nabla_M A_\mu\,^M) 
\eeq
and gauge field weight
\beq \label{15}
\de (A_{\sl 3}\,^M)\cdot \exp\,i\,\int \left\{{\cal L}_{ID} + \ep \mbox{-terms} \right\},
\eeq
where
\beq \label{16}
{\cal L}_{ID} = -\frac{\La^2}{4} \, F_{\mu\nu}\,^M \cdot F^{\mu\nu}\,_M \eeq
is the Lagrangian density of Isometrodynamics and
\beq \label{17}
F_{\mu\nu}\,^M = \pa_\mu A_\nu\,^M - \pa_\nu A_\mu\,^M + A_\mu\,^N \cdot \nabla_N A_\nu\,^M - A_\nu\,^N \cdot \nabla_N A_\mu\,^M \eeq
are the covariant field strength components \cite{chw2}. The $\ep$-terms indicate the appropriate imaginary parts of propagators.

Note that the measure Eqn.(\ref{14}) is the gauge-invariant functional measure on the space of gauge fields living in the gauge algebra ${\overline{\bf diff}}\,{\bf R}^D$.

\section{General Gauge Fixing in the De Witt-Faddeev-Popov Approach and Ghosts}
In this section we derive Quantum Isometrodynamics in general gauges based on the De Witt-Faddeev-Popov method. We then introduce the ghost fields related to these gauges and the generating functional for Green functions.

Following closely \cite{stw2} we start noting that gauge invariant Green functions calculated as path integrals with measure and weight given by Eqns. (\ref{14}) and (\ref{15}) respectively are of the general form
\beq \label{18}
{\cal J} = \int\, \Pi_{\!\!\!\!\!\!_{_{_{x,X,n}}}}\!\!\!\!d \phi_n \cdot
{\cal G} \left[\phi \right] B \left[ f[\phi] \right]
\Det\, {\cal F} \left[\phi \right],
\eeq
where $\phi_n (x,X)$ are a set of gauge and matter fields, $\Pi_{\!\!\!\!\!\!_{_{_{x,X,n}}}}\!\!\!\!d \phi_n$ is a volume element and ${\cal G} \left[\phi \right]$ is a functional of the $\phi_n$ satisfying the gauge-invariance requirement
\beq \label{19}
\Pi_{\!\!\!\!\!\!_{_{_{x,X,n}}}}\!\!\!\!d \phi_{_{\cal E}\,n} \cdot
{\cal G} \left[\phi_{_{\cal E}} \right] =^{\!\!\!\!{!}}\:
\Pi_{\!\!\!\!\!\!_{_{_{x,X,n}}}}\!\!\!\!d \phi_n \cdot
{\cal G} \left[\phi \right].
\eeq
$\phi_{_{\cal E}\,n}$ denote the fields after an infinitesimal gauge transformation with local gauge parameters ${\cal E}^M (x,X)$, $f_R [\phi;x,X]$ is a vector-valued non gauge invariant "gauge fixing functional", $B \left[ f \right]$ a numerical functional defined on general $f$ and ${\cal F}$ is the operator
\beq \label{20}
{\cal F}^R\,_S \left[\phi \right](x,X) \equiv \frac{\de f^R [\phi_{_{\cal E}}](x,X)}{\de\, {\cal E}^S (x,X)} _{\mid_{_{{\cal E}=0}}}. \eeq
Indeed, with fields $\phi_n$ taken as $A_\mu\,^M$ and $\psi_m$, and setting
\beqq \label{21}
f^R [A,\psi] &=& A_{\sl 3}\,\!^R, \nonumber \\
B \left[ f \right] &=& \Pi_{\!\!\!\!\!\!_{_{_{x,X,R}}}}
\!\!\!\de\left(f^R (x,X) \right), \nonumber \\
{\cal G} [A,\psi] &=& \exp\,i\,\int \left\{{\cal L}_{ID} + {\cal L}_M +  \ep \mbox{-terms} \right\} \\
&\times & \mbox{gauge invariant functionals of $A,\psi$} \nonumber \\
\Pi_{\!\!\!\!\!\!_{_{_{x,X,n}}}}\!\!\!\!d \phi_n &=&
\Pi_{\!\!\!\!\!\!_{_{_{x,X,m}}}}
\!\!\!\!d\psi_m \cdot
\Pi_{\!\!\!\!\!\!_{_{_{x,X,M,\mu}}}}
\!\!\!\!\!\!\!\!dA_\mu\,^M \;
\Pi_{\!\!\!\!\!_{_{_{\mu}}}} \,\, \de (\nabla_M A_\mu\,^M)
\nonumber \eeqq
the integral ${\cal J}$ Eqn.(\ref{18}) yields the Green functions of Isometrodynamics in the Euclidean plus axial gauge as defined above. Here we have used the fact that
\beq \label{22} {\cal F}^R\,_S [A_\mu\,^M](x,X) =
\de^R\,_S\cdot \pa_{\sl 3} \eeq
is field-independent and $\Det\, {\cal F}$ reduces to an overall normalization factor in the Euclidean plus axial gauge.

Next, let us check the gauge-invariance requirement Eqn.(\ref{19}). Under local gauge transformations we have
\beqq \label{23}
& & A_{_{\cal E}\,\mu}\,^M = A_\mu\,^M + \pa_\mu {\cal E}^M + A_\mu\,^N \cdot \nabla_N {\cal E}^M - {\cal E}^N \cdot \nabla_N A_\mu\,^M, \nonumber \\
& & \Pi_{\!\!\!\!\!\!_{_{_{x,X,M,\mu}}}}\!\!\!\!\!\!\!\!dA_{_{\cal E}\,\mu}\,^M = \Det \left( \frac{\de A_{_{\cal E}\,\mu}\,^M }{\de A_\nu\,^N } \right) \cdot \Pi_{\!\!\!\!\!\!_{_{_{x,X,M,\mu}}}}\!\!\!\!\!\!\!\!dA_\mu\,^M, \\
& & \de (\nabla_M A_{_{\cal E}\,\mu}\,^M) = \de (\nabla_M A_\mu\,^M). \nonumber
\eeqq
Calculating
\beq \label{24}
\frac{\de A_{_{\cal E}\,\mu}\,^M }{\de A_\nu\,^N } = \eta_\mu\,^\nu\cdot\left( \de^M\,\!_N + \nabla_N {\cal E}^M - {\cal E}^K\cdot \nabla_K\, \de^M\,\!_N \right) 
\eeq
we find that the functional trace of the logarithm of the above Jacobian vanishes - yielding $\Det (\dots)=1$ in Eqn.(\ref{24}). As a result the gauge field measure is gauge invariant and the condition Eqn.(\ref{19}) is fulfilled.

Now we are in a position to freely change the gauge as path integrals of the form Eqn.(\ref{18}) are actually independent of the gauge-fixing functional $f^R [\phi;x,X]$ and depend on the choice of the functional $B \left[ f \right]$ only through an irrelevant constant. The proof of this crucial theorem is found e.g. in \cite{stw2} - as all the steps in the proof hold true for Isometrodynamics as well we do not repeat them explicitly here.

As a result the generating functional for the Green functions of QID in an arbitrary gauge and in the presence of "matter" fields is given by
\beqq \label{25} {\cal Z}\left[\eta, J \right]
&\equiv& \int\Pi_{\!\!\!\!\!\!_{_{_{x,X,m}}}}
\!\!\!\!d\psi_m \cdot
\int\Pi_{\!\!\!\!\!\!_{_{_{x,X,M,\mu}}}}
\!\!\!\!\!\!\!\!dA_\mu\,^M \;
\Pi_{\!\!\!\!\!_{_{_{\mu}}}} \,\,
\de (\nabla_M A_\mu\,^M) 
\\
& & \cdot \exp\,i\,\left\{S_{ID} + S_M + \La^2\, \int J\cdot A + \int \sum_m \eta_m\cdot \psi_m + \ep \mbox{-terms} \right\} \nonumber \\
& & \cdot B \left[ f [A,\psi] \right]
\Det\, {\cal F} \left[A,\psi \right], \nonumber \eeqq
where we have introduced the external sources $\eta$ and $J$ - transforming as a vector in "inner" space - for the "matter" and gauge fields respectively.

In order to further evaluate the generating functional above we choose
\beqq \label{26}
B\left[f [A,\psi] \right]&\equiv& \exp \,i \, S_{GF} \nonumber \\
S_{GF} &\equiv& - \frac{\La^2}{2\xi}\,\intx\intX \La^D \,
f_R [A,\psi] \cdot f^R [A,\psi]
\eeqq
to be quadratic in the gauge-fixing functional $f^R [A,\psi]$ which transforms as a vector in "inner" space and reexpress the functional determinant as the Gaussian integral
\beqq \label{27}
\Det\, {\cal F} \left[A,\psi \right] &\propto&
\int\,\Pi_{\!\!\!\!\!\!_{_{_{x,X,R}}}}\!\!d\om^*_R \;
\de (\nabla^R \om^*_R) \cdot \int\,\Pi_{\!\!\!\!\!\!_{_{_{x,X,S}}}}\!\!d\om^S \;
\de (\nabla_S\, \om^S) \cdot \exp \,i \, S_{GH} \nonumber \\
S_{GH} &\equiv& \La^2 \, \intx\intX \La^D \, \om^*_R \cdot
{\cal F}^R\,_S \left[A,\psi \right] \om^S.
\eeqq
Above we have introduced the ghost fields $\om^*_R (x,X)$ and $\om^S (x,X)$ which are independent anticommuting classical variables. The $\de$-functionals ensure that both sets of variables obey the same constraints as the gauge parameters ${\cal E}$ and that the corresponding operators $\om^* \equiv \om^*_R \nabla^R$ and $\om \equiv \om^S \nabla\!_{S\,}$ are elements of the gauge algebra ${\overline{\bf diff}}\,{\bf R}^D$ which proves crucial in defining the BRST-symmetry operation later.

What is the condition to represent $\Det\, {\cal F} \left[A,\psi \right]$ above as a Gaussian integral as in Eqn.(\ref{27})?

The condition is that for $\om^S$ in the gauge algebra ${\cal F}^R\,_S\, \om^S$ is in the gauge algebra as well. Then
\beq \label{28}
{\cal F}^R\,_S:\:{\overline{\bf diff}}\,{\bf R}^D\: \lar \:\:{\overline{\bf diff}}\,{\bf R}^D
\eeq
is an endomorphism of ${\overline{\bf diff}}\,{\bf R}^D$. Defining the scalar product
\beq \label{29}
\langle g \!\mid\! h \rangle \equiv \La^2 \, \intX \La^D g_M^\dagger (x,X) \cdot h^M (x,X)
\eeq
on ${\overline{\bf diff}}\,{\bf R}^D$ and restricting ourselves to vector-valued functions in ${\overline{\bf diff}}\,{\bf R}^D$ which are square-integrable in the sense of the scalar product above the corresponding function space becomes a Hilbert space. For ${\cal F}^R\,_S$ a selfadjoint endomorphism of ${\overline{\bf diff}}\,{\bf R}^D$ with a complete system of orthonormal eigenvectors we indeed have Eqn.(\ref{27}) with the $\de$-functionals automatically taken account of in the Gaussian integration.

Finally we can write the generating functional for the Green functions of QID in an arbitrary gauge as
\beqq \label{30}
{\cal Z}\left[\eta, J \right]
&\equiv& \int\Pi_{\!\!\!\!\!\!_{_{_{x,X,m}}}}
\!\!\!\!d\psi_m \cdot
\int\Pi_{\!\!\!\!\!\!_{_{_{x,X,M,\mu}}}}
\!\!\!\!\!\!\!\!dA_\mu\,^M \;
\Pi_{\!\!\!\!\!_{_{_{\mu}}}} \,\,
\de (\nabla_M A_\mu\,^M) \nonumber \\
& & \!\!\!\!\!\!\!\!\!\!\!\!
\cdot \int\,\Pi_{\!\!\!\!\!\!_{_{_{x,X,R}}}}\!\!d\om^*_R \;
\de (\nabla^R \om^*_R)
\cdot \int\,\Pi_{\!\!\!\!\!\!_{_{_{x,X,S}}}}\!\!d\om^S \;
\de (\nabla_S\, \om^S) \\
& & \!\!\!\!\!\!\!\!\!\!\!\!
\cdot \exp\,i\,\left\{S_{MOD} + S_M + \La^2\, \int J\cdot A + \int \sum_m \eta_m\cdot \psi_m + \ep \mbox{-terms} \right\}, \nonumber  \eeqq
where
\beq \label{31}
S_{MOD}\equiv S_{ID} + S_{GF} + S_{GH} \eeq
is the modified FP gauge-fixed action for Isometrodynamics.

Eqn.(\ref{30}) defines QID and is the starting point for the evaluation of matrix elements at the quantum level.

\section{Perturbative Expansion, Feynman Rules and Asymptotic States}
In this section we derive the perturbative expansion of the generating functional for the Green functions of pure QID and its Feynman rules in Lorentz-covariant Euclidean gauges. We then use power counting to demonstrate the superficial renormalizability of QID. Finally we analyze the asymptotic states of the theory and are led to introduce additional quantum numbers related to the "inner" degrees of freedom of QID.

Working in Euclidean gauges with "inner" metric $\de_{MN}$ we use Eqn.(\ref{30}) as the starting point for perturbation theory. Splitting the action
\beq \label{32}
S_{MOD} [A, \om^*, \om] \equiv S_{\sl 0} [A, \om^*, \om]
+ S_{INT} [A, \om^*, \om] \eeq
into the part $S_{\sl 0}$ quadratic in the gauge and ghost fields and the interaction part $S_{INT}$ we can rewrite Eqn.(\ref{30}) for pure Isometrodynamics as
\beq \label{33}
{\cal Z}\left[J, \ze^*, \ze \right] = \exp\,i\, S_{INT} \left[\frac{\de\rvec }{\de J}, \frac{\de\rvec}{\de \ze} ,\frac{\de\rvec}{\de \ze^*}\right]  \, {\cal Z}_{\sl 0}\left[J, \ze^*, \ze \right],
\eeq
where
\beqq \label{34}
{\cal Z}_{\sl 0}\left[J, \ze^*, \ze \right]
&\equiv&
\int\Pi_{\!\!\!\!\!\!_{_{_{x,X,M,\mu}}}}
\!\!\!\!\!\!\!\!dA_\mu\,^M \;
\Pi_{\!\!\!\!\!_{_{_{\mu}}}} \,\,
\de (\nabla_M A_\mu\,^M) \nonumber \\
& & \cdot \int\,\Pi_{\!\!\!\!\!\!_{_{_{x,X,R}}}}\!\!d\om^*_R \;
\de (\nabla^R \om^*_R)
\cdot \int\,\Pi_{\!\!\!\!\!\!_{_{_{x,X,S}}}}\!\!d\om^S \;
\de (\nabla_S\, \om^S) \\
& & \cdot \exp\,i\,\left\{S_{\sl 0} + \La^2\, \int (J\cdot A + \om^*\cdot \ze + \ze^*\cdot \om) + \ep \mbox{-terms} \right\} \nonumber \eeqq
is the the generating functional for Green functions of the non-interacting theory and $\ze$, $\ze^*$ are sources for the ghost fields. Note that for consistency reasons all $J$, $\ze$, $\ze^*$ have to be elements of the gauge algebra ${\overline{\bf diff}}\,{\bf R}^D$. In particular, it is crucial that the conserved gauge-field currents
\beq \label{35} J_\nu\,^M = A^{\mu\,N}\cdot \nabla_N F_{\mu\nu}\,^M
- F_{\mu\nu}\,^N\cdot \nabla_N A^{\mu\,M}
\eeq
related to the global coordinate transformation invariance in "inner" space and generating the self coupling of the gauge fields are elements of the gauge algebra ${\overline{\bf diff}}\,{\bf R}^D$ which is easily verified.

To derive Feynman rules we have to specify the gauge and choose
\beq \label{36}
f^R [A] \equiv \pa^\mu A_\mu\,^R
\eeq
as the Lorentz-covariant gauge fixing function resulting in
\beq \label{37}
{\cal F}^R\,_S \left[A\right] = \pa^\mu\! \left(\pa_\mu \,\de^R\,\!_S + A_\mu\,^K\cdot \nabla_K \,\de^R\,\!_S - \nabla_S A_\mu\,^R\right) 
\eeq
which is easily shown to be an endomorphism of ${\overline{\bf diff}}\,{\bf R}^D$ as required.

For the choice Eqn.(\ref{36}) $S_{\sl 0}$ is calculated to be
\beqq \label{38}
S_{\sl 0} &=& - \frac{\La^2}{2} \, \int \La^{D} \, A_\mu\,^M \cdot
{\cal D}_{{\sl 0},\xi}^{\mu\nu}\,_{MN}\, A_\nu\,^N \nonumber \\
&-& \,\La^2 \, \int \La^D \, \om^*_R \cdot {\cal D}_{\sl 0}^R\,_S\, \om^S,
\eeqq
where we have defined the non-interacting gauge and ghost field fluctuation operators by
\beqq \label{39}
{\cal D}_{{\sl 0},\xi}^{\mu\nu}\,_{MN} &\equiv& \left( -\, \eta ^{\mu\nu} \cdot \pa^2  + \left(1 - \frac{1}{\xi} \right)\,
\pa^\mu \, \pa^\nu \right) \de_{MN} \nonumber \\
{\cal D}_{\sl 0}^R\,_S &\equiv& -\, \pa^2\, \de^R\,_S
\eeqq
and the corresponding free propagators $G^{\sl 0}$ through
\beqq \label{40}
{\cal D}_{{\sl 0},\xi}^{\mu\rho}\,_{MR} \,
G^{{\sl 0},\xi}_{\rho\nu}\,^{RN} (x,y; X,Y) &=& \de_M\,^N \, \La^{-D}\, \de^D (X-Y) \, \eta^\mu\,_\nu \, \de^4 (x-y) \nonumber \\
{\cal D}_{\sl 0}^R\,_M \, G_{\sl 0}^M\,_S (x,y; X,Y) &=& \de^R\,_S \, \La^{-D}\, \de^D (X-Y) \, \de^4 (x-y).
\eeqq
The factors of $\La$ ensure the scale invariance of the r.h.s under "inner" scale transformations. After some algebra we find the propagators for the gauge and ghost fields to be
\beqq \label{41}
& & G^{{\sl 0},\xi}_{\mu\nu}\,^{MN} (x,y; X,Y) =
\La^{-D}\, \de^D (X-Y)\, \de^{MN}\cdot 
\nonumber \\
& & \quad\quad\quad\quad\quad
\cdot \int\! \frac{d^4 k}{(2{\pi})^4}\, e^{i\,k\cdot (x-y)}
\, \frac{1}{k^2 - i\,\ep} \left(\eta_{\mu\nu} -
(1-\xi) \frac{k_\mu k_\nu}{k^2} \right) \\
& & G_{\sl 0}^R\,_S (x,y; X,Y) = 
\La^{-D}\, \de^D (X-Y)\, \de^R\,_S\cdot 
\nonumber \\
& & \quad\quad\quad\quad\quad
\cdot \int\! \frac{d^4 k}{(2{\pi})^4}\, e^{i\,k\cdot (x-y)}
\, \frac{1}{k^2 - i\,\ep}. \nonumber \eeqq
They are manifestly diagonal and local in "inner" space and invariant under local Euclidean transformations $X^M\ar X'^M = A^M (x) + O^M\,_N (x)\, X^N$, $O^M\,_N\in SO(D)$. The factors of $\La$ naturally ensure that the integration measure in "inner" $K$-space is dimensionless.

Both the fluctuation operators and the propagators are endomorphisms of ${\overline{\bf diff}}\,{\bf R}^D$, i.e. if $f^M$ fullfills $\nabla_M f^M =0$ so will ${\cal D}_{{\sl 0},\xi}^{\mu\nu}\,_{MN} f^N$, $G^{{\sl 0},\xi}_{\mu\nu}\,^{MN} f_N$ and ${\cal D}_{\sl 0}^R\,_S f^S$, $G_{\sl 0}^R\,_S f^S$ as is easily verified. In other words the propagators are the inverses of the fluctuation operators on the functional space ${\overline{\bf diff}}\,{\bf R}^D$. As a consequence the $\de$-functions in the measure in Eqn.(\ref{34}) will be automatically taken care of in the Gaussian integrals above.

Performing the Gaussian integrals over the gauge and ghost fields we find
\beqq \label{42}
{\cal Z}_{\sl 0}\left[J, \ze^*, \ze \right]
&\propto&
\exp\,i\, \frac{\La^2}{2} \int\!\!\int J^\mu\,_M\cdot
G^{{\sl 0},\xi}_{\mu\nu}\,^{MN} \, J^\nu\,_N \nonumber\\
& &
\cdot \exp\,i\, \La^2 \int\!\!\int \ze^*_R\cdot
G_{\sl 0}^R\,_S \, \ze^S
\eeqq
up to the functional determinants of the fluctuation operators Eqns.(\ref{39}). These field-independent normalization factors do not contribute to physical amplitudes and can be discarded.

Insertion of the result above into Eqn.(\ref{33}) gives the unrenormalized perturbation expansion of the generating functional of the Green functions of QID which is plagued by the usual ultraviolet and infrared divergencies of perturbative QFT. On top of these we will have to deal with potentially divergent integrals over "inner" space. We will show that they can be consistently defined respecting the "inner" scale invariance of the classical theory.

Next, we give the momentum space Feynman rules which are easily derived generalizing the usual approach by Fourier-transforming "inner" space integrals as well.

The momentum space gauge field and ghost propagators are given by
\beqq \label{43}
G^{{\sl 0},\xi}_{\mu\nu}\,^{MN} (k; K) &=&
\frac{1}{k^2 - i\,\ep} \left(\eta_{\mu\nu} - (1-\xi) \frac{k_\mu k_\nu}{k^2} \right)\, \de^{MN} \nonumber \\
G_{\sl 0}^R\,_S (k; K) &=& \frac{1}{k^2 - i\,\ep}\,\, \de^R\,_S
\eeqq
being unity in "inner" space. The "inner" degrees of freedom do not propagate whereas the spacetime parts of the propagators equal the well-known Yang-Mills propagators.

The particle content is now easily read off - there is an uncountably infinite number of both massless gauge and unphysical ghost fields - the latter to counter-balance the unphysical gauge field degrees of freedom arising in covariant gauges. Note that the positive-definiteness of the Euclidean metric $\de_{MN}$ with signature $D$ is crucial to ensure unitarity of the theory. An indefinite metric in "inner" space would make Isometrodynmics unviable as a physical theory.

Next, we calculate the vertices starting with the tri-linear gauge field self-coupling
\beq \label{44}
-\, \La^2\, \left(\pa_\mu A_\nu\,^M - \pa_\nu A_\mu\,^M\right)\, A^\mu\,_N \cdot \nabla^N A^\nu\,_M
\eeq
corresponding to a vertex with three vector boson lines. If these lines carry incoming spacetime momenta $k_1$, $k_2$, $k_3$, "inner" momentum space coordinates $K_1$, $K_2$, $K_3$ and gauge field indices $\mu M$, $\nu N$, $\la L$ the contribution of such a vertex to a Feynman integral is
\beqq \label{45}
& & -\, 2\, \La^2\, \Big\{ K_1^L\,\de^{MN}\,( k_{2\,\la} \eta_{\mu\nu} - k_{2\,\mu} \eta_{\nu\la}) \nonumber \\
& &\quad\quad +\,\, K_2^M\,\de^{NL}\,( k_{3\,\mu} \eta_{\nu\la} - k_{3\,\nu} \eta_{\la\mu}) \\
& &\quad\quad +\,\, K_3^N\,\de^{LM}\,( k_{1\,\nu} \eta_{\la\mu} - k_{1\,\la} \eta_{\mu\nu}) \Big\} \nonumber 
\eeqq
with
\beq \label{46}
k_1+k_2+k_3=0,\quad\quad K_1+K_2+K_3=0.
\eeq

The quadri-linear gauge field self-coupling term
\beq \label{47}
-\, \frac{\La^2}{2}\, \left(A_\mu\,^N \cdot \nabla_N A_\nu\,^M - A_\nu\,^N \cdot \nabla_N A_\mu\,^M\right)\, A^\mu\,_R \cdot \nabla^R A^\nu\,_M
\eeq
corresponds to a vertex with four vector boson lines. If these lines carry incoming spacetime momenta $k_1$, $k_2$, $k_3$, $k_4$, "inner" momentum space coordinates $K_1$, $K_2$, $K_3$, $K_4$ and gauge field indices $\mu M$, $\nu N$, $\rho R$, $\si S$ the contribution of such a vertex to a Feynman integral is
\beqq \label{48}
& & -\, \La^2\, \Big\{( K_1^R\, K_2^S\, \de^{MN} - K_2^S\, K_3^M\, \de^{NR}
+ K_3^M\, K_4^N\, \de^{RS} - K_1^R\, K_4^N\, \de^{MS}) \nonumber \\
& & \quad\quad\quad\quad\quad\quad\quad\quad\quad\quad\quad \cdot( \eta_{\mu\nu}\eta_{\rho\si} - \eta_{\mu\si}\eta_{\nu\rho}) \nonumber \\
& & \quad\quad +\,\, ( K_1^S\, K_2^R\, \de^{MN} - K_1^S\, K_3^N\, \de^{MR}
+ K_3^N\, K_4^M\, \de^{RS} - K_2^R\, K_4^M\, \de^{NS}) \nonumber \\
& & \quad\quad\quad\quad\quad\quad\quad\quad\quad\quad\quad \cdot( \eta_{\mu\nu}\eta_{\rho\si} - \eta_{\mu\rho}\eta_{\nu\si}) \\
& & \quad\quad +\,\, ( K_1^N\, K_3^S\, \de^{MR} - K_1^N\, K_4^R\, \de^{MS}
+ K_2^M\, K_4^R\, \de^{NS} - K_2^M\, K_3^S\, \de^{NR}) \nonumber \\
& & \quad\quad\quad\quad\quad\quad\quad\quad\quad\quad\quad \cdot( \eta_{\mu\rho}\eta_{\nu\si} - \eta_{\mu\si}\eta_{\nu\rho})\Big\} \nonumber
\eeqq
with
\beq \label{49}
k_1+k_2+k_3+k_4 = 0,\quad\quad K_1+K_2+K_3+K_4 = 0.
\eeq

Finally, the gauge-ghost field coupling term
\beq \label{50}
-\, \La^2\, \pa^\mu \om^*_R \left(A_\mu\,^S\cdot \nabla_S \,\om^R -\, \om^S\cdot \nabla_S A_\mu\,^R\right)
\eeq
corresponds to a vertex with one outgoing and one incoming ghost line as well as one vector boson line. If these lines carry incoming spacetime momenta $k_1$, $k_2$, $k_3$, "inner" momentum space coordinates $K_1$, $K_2$, $K_3$ and field indices $R$, $S$, $\mu M$ the contribution of such a vertex to a Feynman integral becomes
\beq \label{51}
-\, \La^2\, ( K_2^M\,\de^{RS} -\, K_3^S\,\de^{MR})\, k_{1\,\mu}
\eeq
with
\beq \label{52}
k_1+k_2+k_3=0,\quad\quad K_1+K_2+K_3=0.
\eeq

In summmary, the above propagators and vertices allow us to evaluate the Green functions of QID perturbatively. Note that for any Feynman graph the analogue of the sums over Lie algebra structure constants in Yang-Mills theories are integrals over "inner" momentum space variables with the scale-invariant measure
\beq \label{53}
\int\! \frac{d^D K}{(2{\pi})^D}\,\La^{-D}.
\eeq
As the vertices in such graphs contribute polynomials in the "inner" space coordinates $K_M$ to the integrand and as these "inner" degrees of freedom do not propagate such integrals look badly divergent – we will show in the next section that they can be consistently defined respecting the "inner" scale invariance of the classical theory.

Turning to the spacetime integrals and renormalizability in the power-counting sense we note that the gauge and ghost fields have the same canonical dimensions $[A]=1$ and $[\om^*]=[\om]=1$ relevant for power counting as do their Yang-Mills counterparts. 

The corresponding divergence indices $\de_1$ of the tri-linear gauge field vertex, $\de_2$ of the quadri-linear gauge field vertex and $\de_3$ of the ghost-gauge field vertex vanish
\beq \label{54}
\de_1 = \de_3 = b + d - 4 = 3 + 1 - 4 = 0,\quad\quad
\de_2 = 4 - 4 = 0,
\eeq
where $b$ is the number of gauge field and ghost lines and $d$ the number of spacetime derivatives attached to the respective vertex. Accordingly the superficial degree of divergence $\om$ for any diagram with a total of $B$ external gauge field and ghost lines becomes
\beq \label{55}
\om = 4 - B
\eeq 
which shows that only a finite number of combination of external lines will yield divergent integrals. As a result Isometrodynamics is renormalizable by power counting.

Let us finally consider the classification of asymptotic one-particle states assuming they exist and are not confined which will be further analyzed in the next section.

To label the physical state-vectors we construct a basis of the one-particle Hilbert space of QID given by simultaneous eigenvectors of observables commuting amongst themselves as well as with the Hamiltonian of the theory. In other words we look for a complete system of conserved, commuting observables for QID.

The specific difference of Isometrodynamics to a Yang-Mills theory arises from the structure of the gauge group - all observables not related to the gauge group remain the same and comprise the energy, the momentum and angular momentum three-vectors and other conserved internal degrees of freedom.

As Isometrodynamics and "matter" field Lagrangians minimally coupled to QID are translation and rotation invariant in "inner" space, we have the corresponding conserved observables - the "inner" space "momentum" operator ${\bf K\/}_M$ and the "inner" angular momentum tensor. As ${\bf K\/}_M$ commutes with the already identified set of observables including the Hamiltonian (consistent with the Coleman-Mandula theorem) its eigenvalues $K_M$ become additional quantum numbers labelling physical states. In addition, as all "matter" fields transform as scalars and the gauge and ghost fields as vectors under "inner" rotations "inner" spin becomes yet another quantum number.

As a result we can find a basis of the one-particle Hilbert space 
\beq \label{56}
\mid k_\mu, \si; K_M, \Si; \mbox{all other quantum numbers}\rangle
\eeq
labeled by the momentum four-vector $k_\mu$, the spin $\si$, the "inner" space momentum vector $K_M$ and the "inner" spin $\Si$ which is $0$ for "matter" and $1$ for the gauge and ghost fields of QID.

\section{Effective Action, Renormalization at One- Loop and Asymptotic Freedom}
In this section starting from the formal perturbative expansion derived in the last section we calculate the renormalized effective action at one loop. The crucial point is to note that spacetime and "inner" space integrals in the calculation of loop graphs completely decouple which allows us to first {\it define} the potentially divergent "inner" space integrals appropriately. Note that any consistent definition must respect the "inner" scale invariance of the classical action at the quantum level as this linearly realized symmetry is a symmetry of the quantum effective action as well \cite{stw2}. This allows us second to deal in the usual way with the ultraviolet divergencies related to the short distance behaviour in spacetime and demonstrate the renormalizability of QID at one loop.

Technically we derive a formal expression for the one-loop effective action of pure Isometrodynamics working in a covariant Euclidean background field gauge. We then define the "inner" momentum integrals using $\La$ as a cut-off and demonstrate the locality of the one-loop effective action in "inner" space. To prove the renormalizability at one loop we calculate the divergent contributions to the functional determinant of a general fluctuation operator with differential operator-valued coefficients in four spacetime dimensions. Finally we determine the one-loop counterterms, renormalize the one-loop effective action and calculate the $\beta$-function of both pure Isometrodynamics and QID coupled to the Standard Model fields.

\subsection{Formal Expression}
To derive a formal expression for the one-loop effective action of Isometrodynamics we work in a covariant Euclidean background field gauge choosing
\beq \label{57}
f^R [A,B] \equiv {_B\!{\cal D}}_\mu^R\,\!_S A^{\mu S} \eeq
where
\beq \label{58}
{_B\!{\cal D}}_\mu^R\,\!_S
\equiv  \pa_\mu \,\de^R\,\!_S + B_\mu\,^K\cdot \nabla_K \,\de^R\,\!_S - \nabla_S B_\mu\,^R \eeq
is the covariant derivative in the presence of a background field $B$ we will further specify below.

To get the one-loop expression for the generating functional Eqn.(\ref{30}) we have to expand the exponent around its stationary point up to second order in the fluctuations. Starting with
\beqq \label{59}
S_{MOD} [A, \om^*, \om; B] &=& -\frac{\La^2}{4 g^2} \,\int \La^D \,
F_{\mu\nu}\,^M \cdot F^{\mu\nu}\,_M \nonumber \\
&-& \frac{\La^2}{2\xi g^2} \,\int \La^D \,
{_B\!{\cal D}}^\mu_{R M} A_\mu\,^M \cdot {_B\!{\cal D}}_\nu^R\,\!_N A^{\nu N} \\
&+& \La^2 \, \int \La^D \, \om^*_R \cdot {\cal F}^R\,_S \left[A,B \right] \om^S, \nonumber
\eeqq
where we have explicitly introduced a dimensionless gauge coupling $g^2$ and where
\beqq \label{60}
{\cal F}^R\,_S \left[A, B \right] &=&
\frac{\de}{\de\, {\cal E}^S} \,
{_B\!{\cal D}}_\mu^R\,\!_M A_{_{\cal E}}^{\mu M}\,_{\mid_{_{{\cal E}=0}}}
\nonumber \\
&=& {_B\!{\cal D}}_\mu^R\,\!_M \, {\cal D}^{\mu M}\,\!_S \eeqq
is easily shown to be an endomorphism of ${\overline{\bf diff}}\,{\bf R}^D$ as required, we get the field equations in the presence of $J$ and $B$
\beqq \label{61}
& & {\cal D}_\mu^N\,\!_M F^{\mu\nu M} 
+ \frac{1}{\xi} \, {_B\!{\cal D}}^{\nu N}\,\!_R \, {_B\!{\cal D}}_\mu^R\,\!_M
A^{\mu M} + g^2\, J^{\nu N} = 0 \nonumber \\
& & {_B\!{\cal D}}_\mu^R\,\!_M \, {\cal D}^{\mu M}\,\!_S \, \om^S = 0 \\
& & {_B\!{\cal D}}^\mu_S\,^M \, {\cal D}_{\mu M}\,\!^R \, \om^*_R = 0.
\nonumber \eeqq
They determine the stationary points $A_\mu\,^M = A_\mu\,^M [J, B]$, $\om^S = 0$ and $\om^*_R = 0$ around which we expand. Setting the background field equal to the stationary point
\beq \label{62}
B_\mu\,^M [J] =^{\!\!\!\!{!}}\: A_\mu\,^M [J, B]
\eeq
determines $B$ as a functional of $J$ at least perturbatively.

Next we calculate the second variation of $S_{MOD}$ at the stationary points
\beqq \label{63}
\de^2 S_{MOD} &=& - \La^2 \, \int \La^{D} \, \de A_\mu\,^M \cdot
{\cal D}_{A,\xi}^{\mu\nu}\,_{MN}\, \de A_\nu\,^N \nonumber \\
&-& 2\,\La^2 \, \int \La^D \, \de \om^*_R \cdot {\cal D}_\om^R\,_S\, \de \om^S,
\eeqq
where we have absorbed the factors of $g$ in $\de A_\mu\,^M$ and calculated the gauge and ghost field fluctuation operators to be
\beqq \label{64}
{\cal D}_{A,\xi}^{\mu\nu}\,_{MN} &\equiv& -\, \eta ^{\mu\nu} \cdot {\cal D}^\rho_M\,^R \, {\cal D}_{\rho RN} \, + \left(1 - \frac{1}{\xi} \right)\,
{\cal D}^\mu_M\,^R \, {\cal D}^\nu_{RN} \nonumber \\
& & -\,2\, F^{\mu\nu}\,_R \nabla^R \cdot \de_{MN} 
+ 2\, \nabla_N \, F^{\mu\nu}\,_M \\
{\cal D}_\om^R\,_S &\equiv& -\, {\cal D}^{\rho RM} \, {\cal D}_{\rho MS}.
\nonumber \eeqq
They are endomorphisms of ${\overline{\bf diff}}\,{\bf R}^D$, i.e. if $f^M$ fullfills $\nabla_M f^M =0$ so will ${\cal D}_{A,\xi}^{\mu\nu}\,_{MN} f^N$ and ${\cal D}_\om^R\,_S f^S$, as is easily verified. Note that we had to commute ${\cal D}^\nu_{RN}$ with ${\cal D}^\mu_M\,^R$ to get the expression above for ${\cal D}_{A,\xi}^{\mu\nu}\,_{MN}$.

Taking all together we finally get
\beqq \label{65}
{\cal Z}_{\sl{1}-loop}\left[J \right]
&=& \int\Pi_{\!\!\!\!\!\!_{_{_{x,X,M,\mu}}}}
\!\!\!\!\!\!\!\!d\,\de A_\mu\,^M \; \Pi_{\!\!\!\!\!_{_{_{\mu}}}} \,\,
\de (\nabla_M \de A_\mu\,^M) \nonumber \\
& & \cdot \int\,\Pi_{\!\!\!\!\!\!_{_{_{x,X,R}}}}\!\!d\,\de\om^*_R \;
\de (\nabla^R \de\om^*_R)
\cdot \int\,\Pi_{\!\!\!\!\!\!_{_{_{x,X,S}}}}\!\!d\,\de\om^S \;
\de (\nabla_S\, \de\om^S) \nonumber \\
& & \cdot \exp\,i\,\left\{S_{MOD} [A, 0, 0; A] + \La^2\, \int J\cdot A \right\} \nonumber \\
& & \cdot \exp \Bigg\{ - \frac{i}{2} \, \La^2 \, \int \La^{D} \, \de A_\mu\,^M \cdot {\cal D}_{A,\xi}^{\mu\nu}\,_{MN}\, \de A_\nu\,^N \\
& & \quad\quad\quad - \, i\,\, \La^2 \, \int \La^D \, \de \om^*_R \cdot {\cal D}_\om^R\,_S\, \de \om^S + \ep \mbox{-terms} \Bigg\} \nonumber \\
&=& \exp\,i\,\left\{S_{MOD} [A, 0, 0; A] + \int \La^2\, J\cdot A \right\} \nonumber \\
& & \cdot \Det^{-1/2}\, {\cal D}_{A,\xi} \cdot \Det\, {\cal D}_\om. \nonumber
\eeqq
As the fluctuation operators are endomorphisms of ${\overline{\bf diff}}\,{\bf R}^D$ the integrals in Eqn.(\ref{65}) are Gaussian and can be performed resulting in the usual determinants. Indeed, endowed with the scalar product Eqn.(\ref{29}), ${\overline{\bf diff}}\,{\bf R}^D$ becomes a Hilbert space with a complete orthonormal set of eigenvectors for each of the selfadjoint fluctuation operators above. These bases of the Hilbert space take the $\de$-functionals automatically into account and the integration over each eigenvector direction becomes Gaussian.

Defining next the generating functional for connected Green functions
\beq \label{66}
{\cal W}\left[J \right] \equiv -\, i\, \Ln \, {\cal Z}\left[J \right]
\eeq
and the quantum effective action as its Legendre transform 
\beq \label{67}
\Ga\left[A \right] \equiv -\int J\cdot A + {\cal W}
\eeq
in the usual way we find
\beq \label{68}
\Ga_{\sl{1}-loop} \left[A \right]
= S_{MOD} [A, 0, 0; A] + \frac{i}{2} \Tr\Ln\, {\cal D}_{A,\xi} 
- \, i\, \Tr\Ln\, {\cal D}_\om
\eeq
which is the formal expression for the one-loop effective action we were looking for.

From now on we work with the specific choice $\xi = 1$ and drop the subscript $\xi$ to keep the calculations below as simple as possible.

\subsection{Finiteness and Locality of "Inner" Space Integrals}
To get a well-defined quantum theory at the one-loop level we have to show that the functional traces in Eqn.(\ref{68}) above evaluated over the "inner" space can be appropriately defined, an issue which does not arise in Yang-Mills theories of compact Lie groups due to the finite volume of the underlying gauge groups.

To define $\Tr\Ln\, {\cal D}_A$ and $\Tr\Ln\, {\cal D}_\om$ and to demonstrate their locality in "inner" space note that both operators are of the form
\beq \label{69} 
{\cal D} = -{\pa\rvec}^2 + {\cal M}_{IJ}\,{\nabla\rvec}^I{\nabla\rvec}^J
+ {\cal N}_{I}\,{\nabla\rvec}^I + {\cal C},
\eeq
where ${\cal M}_{IJ\mid N}^{\,M}, {\cal N}_{I\mid N}^{\,M}, {\cal C}^M\,_N$ are both matrices in "inner" space and matrix-valued differential operators in Minkowski space. This form is very general and can account for non-covariant Euclidean gauges such as the Euclidean Lorentz gauge of Eqn.(\ref{57}) as well, however, for $\xi\neq 1$ the operator would take an even more general form.

Properly normalizing and expanding the logarithm we obtain
\beqq \label{70}
\Tr\Ln \frac{{\cal D}}{{\cal D}_0}
&=& \Tr\Ln\, {\cal D} - \Tr\Ln\, {\cal D}_0 \nonumber \\
&=& \Tr\Ln\left({\bf 1} - \frac{1}{{\pa\rvec}^2}\,\left({\cal M}_{IJ}\,{\nabla\rvec}^I{\nabla\rvec}^J + {\cal N}_{I}\,{\nabla\rvec}^I + {\cal C}\right) \right) \\
&=& \sum_n \frac{(-)^n}{n}\, \Tr\left[\left(- \frac{1}{{\pa\rvec}^2}\right)
\left({\cal M}_{IJ}\,{\nabla\rvec}^I{\nabla\rvec}^J + {\cal N}_{I}\,{\nabla\rvec}^I 
+ {\cal C}\right)\right]^n
\nonumber \\
&=& \sum_n \frac{(-)^n}{n}\,\Ga^{(n)}, \nonumber
\eeqq
where ${\cal D}_0$ is the operator for vanishing fields. Here we have defined the one-loop contribution with $n$ "vertex" insertions
\beqq \label{71}
\Ga^{(n)} &\equiv& \Tr_{\!\!\!\!\!\!\!\!\!_{_{_{x,X}}}}
\left[\left(- \frac{1}{{\pa\rvec}^2}\right)
\left({\cal M}_{IJ}\,{\nabla\rvec}^I{\nabla\rvec}^J + {\cal N}_{I}\,{\nabla\rvec}^I 
+ {\cal C}\right)\right]^n \nonumber \\
&=& \int\! d^{D}X_1\dots\dots\int\! d^{D}X_n
\int\! \frac{d^{D}P_1}{(2{\pi})^D}\dots\dots
\int\! \frac{d^{D}P_n}{(2{\pi})^D} \nonumber \\
& & \!\!\!\!\!\!\!\!\!\!\!\!\!\Tr_{\!\!\!\!\!\!\!_{_{_{x}}}}\: \Bigg\{ \langle X_1 \!\mid \left(- \frac{1}{{\pa\rvec}^2}\right)
\left({\cal M}_{I_1 J_1}\, {\nabla\rvec}^{I_1}{\nabla\rvec}^{J_1} + {\cal N}_{I_1}\, {\nabla\rvec}^{I_1} + {\cal C}\right) \mid\! P_1\rangle\,\cdot \nonumber \\
& & \quad\quad\quad\quad\quad\quad\quad\quad\quad\quad \vdots \nonumber \\
& & \cdot\langle X_n\!\mid \left(- \frac{1}{{\pa\rvec}^2}\right)
\left({\cal M}_{I_n J_n}\, {\nabla\rvec}^{I_n}{\nabla\rvec}^{J_n} + {\cal N}_{I_n}\, {\nabla\rvec}^{I_n} + {\cal C}\right) \mid\! P_n\rangle \Bigg\} \nonumber \\
& & \!\!\!\!\!\!\!\!\!\!\!\cdot \langle P_1\!\mid\! X_2\rangle\,\cdot\dots\cdot\,\langle P_n\!\mid\! X_1\rangle \\
&=& \int\! d^{D}X_1\dots\dots\int\! d^{D}X_n
\int\! \frac{d^{D}P_1}{(2{\pi})^D}\dots\dots
\int\! \frac{d^{D}P_n}{(2{\pi})^D} \nonumber \\
& & \!\!\!\!\!\!\!\!\!\!\!\!\!\Tr_{\!\!\!\!\!\!\!_{_{_{x}}}}\: 
\Bigg\{\left(- \frac{1}{{\pa\rvec}^2}\right)
\left({\cal M}_{I_1 J_1}\, i P_1^{I_1} i P_1^{J_1} +
{\cal N}_{I_1}\, i P_1^{I_1} + {\cal C}\right)_{X_1} \,\cdot \nonumber \\
& & \quad\quad\quad\quad\quad\quad\quad\quad\quad\quad \vdots \nonumber \\
& & \cdot\left(- \frac{1}{{\pa\rvec}^2}\right)
\left({\cal M}_{I_n J_n}\, i P_n^{I_n} i P_n^{J_n} +
{\cal N}_{I_n}\, i P_n^{I_n} + {\cal C} \right)_{X_n} \Bigg\} \nonumber \\
& & \!\!\!\!\!\!\!\!\!\!\!\cdot \exp \left(i P_1 (X_1 - X_2)
\,+ \dots +\, i P_n (X_n - X_1)\right) \nonumber 
\eeqq
which is manifestly invariant under local Euclidean transformations $X^M\ar X'^M = A^M (x) + O^M\,_N (x) X^N$, $O^M\,_N\in SO(D)$. Above we have inserted $n$ complete systems of both $X$- and $P$-vectors
\bed
{\bf 1} = \intX \mid\! X\rangle\langle X\!\mid, \quad\quad
{\bf 1} = \intP \mid\! P\rangle\langle P\!\mid
\eed
and used $\langle X\!\mid\! P\rangle = \exp(i\, P\cdot X)$ in Cartesian coordinates. Defining new variables
\beqq \label{72}
K_1 &\equiv& P_1 - P_n \nonumber \\
K_2 &\equiv& P_2 - P_1, \quad\quad\quad P_2 = K_2 + P_1 \nonumber \\
&\vdots& \quad\quad\quad\quad\quad\quad\quad\quad\:\: \vdots \\
K_{n-1} &\equiv& P_{n-1} - P_{n-2}, \quad P_{n-1}
= K_{n-1} + \dots + K_2 + P_1 \nonumber \\
K_n &\equiv& P_n - P_{n-1}, \quad\quad P_n
= K_n + \dots + K_2 + P_1 \nonumber
\eeqq
it becomes obvious that the definition above of the $P_1$-integrals over polynomials in $P_1$ requires care in order to avoid potential infinities related to the non-compactness of the gauge group.

In generalization of our approach to defining the classical action of Isometrodynamics we {\it define} such integrals using the cut-off $\La$ introduced in \cite{chw2}
\beqq \label{73}
\Ga^{(n)}_{\La} &\equiv& \int\! d^{D}X_1\dots\dots\int\! d^{D}X_n
\int_{\mid P_1\mid \leq \La} \! \frac{d^{D}P_1}{(2{\pi})^D}
\int\! \frac{d^{D}K_2}{(2{\pi})^D} \dots\dots
\int\! \frac{d^{D}K_n}{(2{\pi})^D} \nonumber \\
& & \!\!\!\!\!\!\!\!\!\!\!\!\!\Tr_{\!\!\!\!\!\!\!_{_{_{x}}}}\:
\Bigg\{\left(- \frac{1}{{\pa\rvec}^2}\right)
\left({\cal M}_{I_1 J_1}\, i P_1^{I_1} i P_1^{J_1} +
{\cal N}_{I_1}\, i P_1^{I_1} + {\cal C}\right)_{X_1} \,\cdot \nonumber \\
& & \quad\quad\quad\quad\quad\quad\quad\quad\quad\quad \vdots \\
& & \!\!\!\!\!\!\!\!\!\!\!\!\!\!\!\!\cdot\left(- \frac{1}{{\pa\rvec}^2}\right)
\Big({\cal M}_{I_n J_n}\,(i P_1^{I_n} + i K_2^{I_n} + \dots + i K_n^{I_n}) (i P_1^{J_n} + i K_2^{J_n} + \dots + i K_n^{J_n}) \nonumber \\
& & \quad\quad\quad +\:\: {\cal N}_{I_n}\, (i P_1^{I_n} + i K_2^{I_n} + \dots + i K_n^{I_n}) + {\cal C} \Big)_{X_n} \Bigg\} \nonumber \\
& & \!\!\!\!\!\!\!\!\!\!\!\cdot \exp \left( -i X_1 (K_2 + \dots + K_n)+ i X_2 K_2 \,+ \dots +\, i X_n K_n \right). \nonumber 
\eeqq

Next, using $i K^L_j \exp(i \sum_{l=2}^n X_l K_l) = {\nabla\rvec}^L_j \exp(i \sum_{l=2}^n X_l K_l)$
and partially integrating we get
\beqq \label{74}
\Ga^{(n)}_\La &=& \int\! d^{D}X_1\dots\dots\int\! d^{D}X_n
\int_{\mid P_1\mid \leq \La} \! \frac{d^{D}P_1}{(2{\pi})^D}
\int\! \frac{d^{D}K_2}{(2{\pi})^D} \dots\dots
\int\! \frac{d^{D}K_n}{(2{\pi})^D} \nonumber \\
& & \!\!\!\!\!\!\!\!\!\!\!\!\!\Tr_{\!\!\!\!\!\!\!_{_{_{x}}}}\:
\Bigg\{\left(- \frac{1}{{\pa\rvec}^2}\right)
\left({\cal M}_{I_1 J_1}\, i P_1^{I_1} i P_1^{J_1} +
{\cal N}_{I_1}\, i P_1^{I_1} + {\cal C}\right)_{X_1} \,\cdot \nonumber \\
& & \quad\quad\quad\quad\quad\quad\quad\quad\quad\quad \vdots \\
& & \!\!\!\!\!\!\!\!\!\!\!\!\!\!\!\!\cdot\,\left(- \frac{1}{{\pa\rvec}^2}\right)
\Big({\cal M}_{I_n J_n}\,
(i P_1^{I_n} - {\nabla\lvec}_2^{I_n} - \dots - {\nabla\lvec}_n^{I_n}) (i P_1^{J_n} - {\nabla\lvec}_2^{J_n} - \dots - {\nabla\lvec}_n^{J_n}) \nonumber \\
& & \quad\quad\quad +\:\: {\cal N}_{I_n}\, (i P_1^{I_n} - {\nabla\lvec}_2^{I_n} - \dots - {\nabla\lvec}_n^{I_n}) + {\cal C} \Big)_{X_n} \Bigg\} \nonumber \\
& & \!\!\!\!\!\!\!\!\!\!\!\cdot \exp \left(i K_2 (X_2 - X_1) \,+ \dots +\, i K_n (X_n - X_1) \right). \nonumber 
\eeqq
Above, the differential operators act to the left and ordering obviously matters. Integrating over $K_i, X_j$ for $i,j=\sl{2,3}\dots n$ yields the final expression for $\Ga^{(n)}_\La $ in this subsection
\beqq \label{75}
\Ga^{(n)}_\La &=& \int\! d^{D}X_1
\int_{\mid P_1\mid \leq \La} \! \frac{d^{D}P_1}{(2{\pi})^D} \nonumber \\
& & \!\!\!\!\!\!\!\!\!\!\!\!\!\Tr_{\!\!\!\!\!\!\!_{_{_{x}}}}\:
\Bigg\{\left(- \frac{1}{{\pa\rvec}^2}\right)
\left({\cal M}_{I_1 J_1}\, i P_1^{I_1} i P_1^{J_1} +
{\cal N}_{I_1}\, i P_1^{I_1} + {\cal C}\right)_{X_1} \,\cdot \nonumber \\
& & \quad\quad\quad\quad\quad\quad\quad\quad\quad\quad \vdots \\
& & \!\!\!\!\!\!\!\!\!\!\!\!\!\!\!\!\cdot\,\left(- \frac{1}{{\pa\rvec}^2}\right)
\Big({\cal M}_{I_n J_n}\,
(i P_1^{I_n} - {\nabla\lvec}_2^{I_n} - \dots - {\nabla\lvec}_n^{I_n}) (i P_1^{J_n} - {\nabla\lvec}_2^{J_n} - \dots - {\nabla\lvec}_n^{J_n}) \nonumber \\
& & +\:\: {\cal N}_{I_n}\, (i P_1^{I_n} - {\nabla\lvec}_2^{I_n} - \dots - {\nabla\lvec}_n^{I_n}) + {\cal C} \Big)_{X_n=X_{n-1}=..=X_1} \Bigg\}. \nonumber
\eeqq
The expression above for $\Ga^{(n)}_\La $ is not only finite as an integral over "inner" space, but also local in $X_1$. Note that the regularized integrals over $P_1$ collapse into sums over products of $\de$-functions in "inner" space. These sums correspond to the sums over structure constants in the Yang-Mills case.

As in the case of the classical Lagrangian the contributions $\Ga^{(n)}$ to the one-loop effective action for $\rho\La$ are related to the ones for a given $\La$ by
\beq \label{76}
\Ga^{(n)}_{\rho\La} (X,A_\nu\,^M(X),\dots) =
\Ga^{(n)}_{\La} (\rho X,\rho A_\nu\,^M(X),\dots)
\eeq
respecting the scale invariance of the classical theory.

At one loop the dependence of the theory on $\La$ is again controlled by the scale invariance of the classical theory. In other words up to one loop theories for different $\La$ are equivalent up to "inner" rescalings. This symmetry is not distroyed by the renormalization required for the divergent spacetime integrals with which we deal in the next subsection for the simple fact that both types of integrals and how we properly define them completely decouple.

\subsection{Divergence Structure of Spacetime Integrals}
We turn to calculate the divergent contributions to the functional determinant of a general fluctuation operator with differential operator-valued coefficients in four spacetime dimensions in preparation of the one-loop renormalization in the next subsection. 

To analyze the divergencies occurring in $\Tr_\La \Ln \, {\cal D}_A$ and $\Tr_\La \Ln\, {\cal D}_\om$ note that both operators are of the form
\beq \label{77} 
{\cal D} = -{\pa\rvec}^2 + {\cal B}_\rho\,{\pa\rvec}^\rho + {\cal C},
\eeq
where ${\cal B}_\rho, {\cal C}$ are both matrices in Minkowski space and matrix-valued differential operators in "inner" space. Again, this form is general enough to cope with non-covariant Euclidean gauges of the form Eqn.(\ref{57}), however, the case $\xi\neq 1$ is not included.

Properly normalizing and expanding the logarithm we obtain
\beqq \label{78}
\Tr_\La \Ln \frac{{\cal D}}{{\cal D}_0}
&=& \Tr_\La \Ln\, {\cal D} - \Tr_\La \Ln\, {\cal D}_0 \nonumber \\
&=& \Tr_\La \Ln\left({\bf 1} - \frac{1}{{\pa\rvec}^2}\,\left({\cal B}_\rho\,{\pa\rvec}^\rho + {\cal C}\right) \right) \\
&=& \sum_n \frac{(-)^n}{n}\, \Tr_\La \left[\left(- \frac{1}{{\pa\rvec}^2}\right)
\left({\cal B}_\rho\,{\pa\rvec}^\rho + {\cal C}\right)\right]^n
\nonumber \\
&=& \sum_n \frac{(-)^n}{n}\,\Ga^{(n)}_\La, \nonumber
\eeqq
where ${\cal D}_0$ is the operator for vanishing fields. Here we have defined
\beqq \label{79}
\Ga^{(n)}_\La &\equiv& {\Tr_{\!\!\!\!\!\!\!\!\!_{_{_{x,X}}}}}_\La 
\left[\left(- \frac{1}{{\pa\rvec}^2}\right)
\left({\cal B}_\rho\,{\pa\rvec}^\rho + {\cal C}\right)\right]^n \nonumber \\
&=& \int\! d^4 x_1\dots\dots\int\! d^4 x_n
\int\! \frac{d^4 p_1}{(2{\pi})^4}\dots\dots
\int\! \frac{d^4 p_n}{(2{\pi})^4} \nonumber \\
& & \!\!\!\!\!{\Tr_{\!\!\!\!\!\!\!_{_{_{X}}}}}_\La \: \Bigg\{ \langle x_1\!\mid \left(- \frac{1}{{\pa\rvec}^2}\right)
\left({\cal B}_{\rho_1}\,{\pa\rvec}_1^{\rho_1} + {\cal C}\right) \mid \!p_1\rangle\,\cdot \nonumber \\
& & \quad\quad\quad\quad\quad\quad\quad \vdots \nonumber \\
& & \quad\quad \cdot\langle x_n\!\mid \left(- \frac{1}{{\pa\rvec}^2}\right)
\left({\cal B}_{\rho_n}\,{\pa\rvec}_n^{\rho_n} + {\cal C}\right) \mid \!p_n\rangle \Bigg\} \nonumber \\
& & \!\!\!\cdot \langle p_1\!\mid\! x_2\rangle\,\cdot\dots\cdot\,\langle p_n\!\mid\! x_1\rangle \\
&=& \int\! d^4 x_1\dots\dots\int\! d^4 x_n
\int\! \frac{d^4 p_1}{(2{\pi})^4}\dots\dots
\int\! \frac{d^4 p_n}{(2{\pi})^4} \nonumber \\
& & \!\!\!\!\!{\Tr_{\!\!\!\!\!\!\!_{_{_{X}}}}}_\La \: \Bigg\{
\frac{1}{p_1^2} \,
\left(i\,{\cal B}_{\rho_1}\,p_1^{\rho_1} + {\cal C}\right)_{x_1} \cdot
\nonumber \\
& & \quad\quad\quad\quad\quad\quad\quad \vdots \nonumber \\
& & \quad\quad \cdot \frac{1}{p_n^2} \,
\left(i\,{\cal B}_{\rho_n}\,p_n^{\rho_n} + {\cal C}\right)_{x_n}\Bigg\} \nonumber \\
& & \!\!\!\cdot \exp \left(i p_1 (x_1 - x_2) \,+ \dots +\, 
i p_n (x_n - x_1)\right), \nonumber 
\eeqq
where we have inserted $n$ complete systems of both $x$- and $p$-vectors
\bed
{\bf 1} = \intx \mid \!x\rangle\langle x\!\mid, \quad\quad
{\bf 1} = \intp \mid \!p\rangle\langle p\!\mid
\eed
and where $\langle x\!\mid\! p\rangle = \exp(i\, p\cdot x)$. Note the occurrence of the propagators above which is in marked difference to the local "inner" space integrals analyzed in the last subsection.

A shift of variables
\beqq \label{80}
k_1 &\equiv& p_1 - p_n \nonumber \\
k_2 &\equiv& p_2 - p_1, \quad\quad\quad p_2 = k_2 + p_1 \nonumber \\
&\vdots& \quad\quad\quad\quad\quad\quad\quad\quad \vdots \\
k_{n-1} &\equiv& p_{n-1} - p_{n-2}, \quad p_{n-1}
= k_{n-1} + \dots + k_2 + p_1 \nonumber \\
k_n &\equiv& p_n - p_{n-1}, \quad\quad p_n
= k_n + \dots + k_2 + p_1 \nonumber
\eeqq
allows us to rewrite $\Ga^{(n)}_\La$ as
\beqq \label{81}
\Ga^{(n)}_\La &=& \int\! d^4 x_1\dots\dots\int\! d^4 x_n
\int\! \frac{d^4 p_1}{(2{\pi})^4}\!
\int\! \frac{d^4 k_2}{(2{\pi})^4} \dots\dots
\int\! \frac{d^4 k_n}{(2{\pi})^4} \nonumber \\
& & \!\!\!\!\!\!\!\!\!\!{\Tr_{\!\!\!\!\!\!\!_{_{_{X}}}}}_\La \: \Bigg\{
\frac{1}{p_1^2} \,
\left(i\,{\cal B}_{\rho_1}\,p_1^{\rho_1} + {\cal C}\right)_{x_1} \cdot
\nonumber \\
& & \quad\quad\quad\quad\quad\quad\quad\quad\quad\quad \vdots \\
& & \!\!\!\!\!\cdot \frac{1}{(p_1 + k_2 + \dots + k_n)^2} \,
\left(i\,{\cal B}_{\rho_n}\,\left(p_1^{\rho_n} + k_2^{\rho_n} + \dots + k_n^{\rho_n}\right) + {\cal C}\right)_{x_n}\Bigg\} \nonumber \\
& & \!\!\!\!\!\!\!\!\cdot \exp \left( -i x_1 (k_2 + \dots + k_n)+ i x_2 k_2 \,+ \dots +\, i x_n k_n \right). \nonumber 
\eeqq
Now it is easy to read off the degrees of divergence $\om_n$ for the $p_1$-integrals which are bound by $\om_n\leq 4-n$. Hence, only the $\Ga^{(n)}_\La$ for $n=1,2,3,4$ have a divergent contribution.

Isolating the divergent contributions which are local in $x_1$ we find
\beqq \label{82}
\left(\Tr_\La \Ln \frac{{\cal D}}{{\cal D}_0}\right)^{div} &=&
\Ga^{(1)\, div}_\La - \frac{1}{2}\, \Ga^{(2)\, div}_\La 
+ \frac{1}{3}\, \Ga^{(3)\, div}_\La - \frac{1}{4}\, \Ga^{(4)\, div}_\La \nonumber \\
&=& i\,\frac{\Om_4}{\ep} \intx_1 \,
{\Tr_{\!\!\!\!\!\!\!_{_{_{X}}}}}_\La \: \Bigg\{
- \frac{1}{12}\, \pa^\mu\,{\cal B}_\mu \cdot \pa^\nu\,{\cal B}_\nu \nonumber \\
&-& \frac{1}{24}\, \pa^\nu\,{\cal B}_\mu \cdot \pa_\nu\,{\cal B}^\mu + \frac{1}{2}\, \pa^\mu\,{\cal B}_\mu \cdot {\cal C}
- \frac{1}{2}\, {\cal C}^2 \\
&+& \frac{1}{12}\, \pa^\mu\,{\cal B}_\mu \cdot {\cal B}^\nu \cdot {\cal B}_\nu 
- \frac{1}{12}\, {\cal B}_\mu \cdot \pa^\mu {\cal B}^\nu \cdot {\cal B}_\nu \nonumber \\
&-& \frac{1}{4}\, {\cal C} \cdot {\cal B}^\nu \cdot {\cal B}_\nu
- \frac{1}{48}\, {\cal B}^\mu \cdot {\cal B}_\mu \cdot {\cal B}^\nu \cdot {\cal B}_\nu \nonumber \\
&-& \frac{1}{96}\, {\cal B}^\mu \cdot {\cal B}^\nu \cdot {\cal B}_\mu \cdot {\cal B}_\nu
\Bigg\}. \nonumber 
\eeqq
Above, we have used the results from Appendix A for the $\Ga^{(n)\, div}_\La$ for $n=1,2,3,4$ with $\ep = d -4$ and $\Om_4 = \frac{1}{8\pi^2}$.

For fluctuations operators of the form
\beq \label{83} 
{\cal D} = -\, {\cal D}_\mu \,{\cal D}^\mu + {\cal E},\quad\quad
{\cal D}_\mu \equiv \pa_\mu + {\cal A}_\mu, 
\eeq
where the gauge field ${\cal A}_\mu$ is a matrix-valued differential operator, we have
\beq \label{84} 
{\cal B}_\mu = -\,2\, {\cal A}_\mu,\quad\quad {\cal C} = -\,\pa_\mu {\cal A}^\mu - {\cal A}_\mu \cdot {\cal A}^\mu + {\cal E}
\eeq
and using the cyclicality property of the trace, which is easily demonstrated, Eqn.(\ref{82}) further simplifies
\beq \label{85} 
\left(\Tr_\La \Ln \frac{{\cal D}}{{\cal D}_0}\right)^{div}
= - i\,\frac{\Om_4}{\ep} \intx_1 \,
{\Tr_{\!\!\!\!\!\!\!_{_{_{X}}}}}_\La \: \Bigg\{
\frac{1}{12}\, {\cal F}_{\mu\nu} \cdot {\cal F}^{\mu\nu} 
+ \frac{1}{2}\, {\cal E}^2 \Bigg\}.
\eeq
Above we have introduced the field strength operator 
\beq \label{86} 
{\cal F}_{\mu\nu} \equiv \left[{\cal D}_\mu, {\cal D}_\nu \right]
\eeq
which belongs to the gauge field operator ${\cal A}_\mu$.

\subsection{One-Loop Renormalization}
With the formulae Eqns.(\ref{85}) and (\ref{86}) which hold true in the presence of general spacetime- and "inner" space-dependent fields we are now in a position to analyze the one-loop renormalizability of Isometrodynamics both in the absence and presence of "matter" fields. Note that after properly regularizing the "inner" space integrals we can safely interchange the order of taking the traces over "inner" space versus spacetime variables. In this section we perform the functional trace over spacetime variables first.

To analyze renormalizability we have to evaluate the divergent contributions to the one-loop effective action $\Ga_{\La, \sl{1}-loop} \left[A \right]$ from Eqn.(\ref{68}). A short calculation shows that the fluctuation operators Eqns.(\ref{64}) take the form of Eqn.(\ref{83}) above with
\beqq \label{87} 
\left({\cal A}_\mu\right)^M\,_N &=&  A_\mu\,^K \nabla_K \,\de^M\,_N
- \nabla_N A_\mu\,^M \nonumber \\
\left({\cal F}_{\mu\nu}\right)^M\,_N &=&  F_{\mu\nu}\,^K \nabla_K \,\de^M\,_N
- \nabla_N F_{\mu\nu}\,^M \\
{\cal D}_A^{\mu\nu}\,_{MN} &=& -\, \eta ^{\mu\nu} \cdot \left({\cal D}^\rho\right)_M\,^R \, \left({\cal D}_\rho\right)_{RN} \, 
- 2\, \left({\cal F}_{\mu\nu}\right)_{MN} \nonumber \\
{\cal D}_\om^R\,_S &=& -\, \left({\cal D}^\mu\right)^{RM} \, \left({\cal D}_\mu\right)_{MS}.
\nonumber \eeqq
As a result we get - after taking the trace over Minkowski indices - the divergent contributions to the gauge field determinant
\beqq \label{88} 
\left(\Tr_\La \Ln \frac{{\cal D}_A}{{\cal D}_0}\right)^{div}
&=& - i\,\frac{\Om_4}{\ep} \intx \,
{\Tr_{\!\!\!\!\!\!\!_{_{_{X}}}}}_\La \: \Bigg\{
\frac{1}{12}\,4\, {\cal F}_{\mu\nu} \cdot {\cal F}^{\mu\nu} 
+ \frac{1}{2}\,4\, {\cal F}_{\mu\nu} \cdot {\cal F}^{\nu\mu} \Bigg\}
\nonumber \\
&=& i\,\frac{\Om_4}{\ep} \,\frac{5}{3}\,D\, \intx 
{\Tr_{\!\!\!\!\!\!\!_{_{_{X}}}}}_\La \: F_{\mu\nu} \cdot F^{\mu\nu},
\eeqq
and to the ghost determinant
\beqq \label{89}
\left(\Tr_\La \Ln \frac{{\cal D}_\om}{{\cal D}_0}\right)^{div}
&=& - i\,\frac{\Om_4}{\ep} \intx \,
{\Tr_{\!\!\!\!\!\!\!_{_{_{X}}}}}_\La \:
\frac{1}{12} \, {\cal F}_{\mu\nu} \cdot {\cal F}^{\mu\nu} \nonumber \\
&=& - i\,\frac{\Om_4}{\ep} \,\frac{1}{12}\,D\, \intx 
{\Tr_{\!\!\!\!\!\!\!_{_{_{X}}}}}_\La \:  F_{\mu\nu} \cdot F^{\mu\nu}.
\eeqq
Note that as for other gauge field theories it is the second term in Eqn.(\ref{88}) which determines the sign of the gauge field contribution above – which will in turn determine the sign of the $\beta$-function of Isometrodynamics.

Taking all together we find
\beqq \label{90}
\Ga_{\La, \sl{1}-loop}^{div} \left[A \right]
&=&
\frac{i}{2} \left(\Tr_\La \Ln \frac{{\cal D}_A}{{\cal D}_0}\right)^{div}
- \, i\, \left(\Tr_\La \Ln \frac{{\cal D}_\om}{{\cal D}_0}\right)^{div} \nonumber \\
&=& 
- \,\frac{\Om_4}{\ep} \,\frac{11}{12}\,D\, \intx {\Tr_{\!\!\!\!\!\!\!_{_{_{X}}}}}_\La \:  F_{\mu\nu} \cdot F^{\mu\nu} \\
&=& \,\frac{\Om_4}{\ep} \,\frac{\Om_D}{D(D+2)} \,\frac{11}{12}\,D 
\,\La^2 \,\int \La^D \,F_{\mu\nu}\,^M \cdot F^{\mu\nu}\,_M.
\nonumber \eeqq
The one-loop divergence is proportional to the action of Isometrodynamics and the theory is renormalizable at one loop. Note the formal similarity of the formula above with the analogous expression for Yang-Mills theories, especially the occurrence of the universal numerical factor $\frac{11}{12}$.

As usual the divergent contribution $\Ga_{\La, \sl{1}-loop}^{div} \left[A \right]$ can be absorbed in the original action of Isometrodynamics through a redefinition of the gauge coupling constant
\beq \label{91}
g_R = g\left(1 + \frac{g^2}{4 \pi^2} \,\frac{\Om_D}{D(D+2)} \,\frac{11}{12}\,D \,\frac{1}{\ep} + O(g^4)\right)
\eeq
where we have used $\Om_4=\frac{1}{8\pi^2}$. 

As a result the one-loop effective action after regularization of the "inner" space integrals and renormalization is a perfectly well defined expression.

The corresponding $\be$-function of Isometrodynamics at one loop becomes
\beq \label{92}
\be(g) = - \frac{g^3}{4 \pi^2} \,\frac{\Om_D}{D(D+2)} \,\frac{11}{12}\,D 
\eeq
and the theory is asymptotically free.

Note that $\La$ does not get renormalized as we would expect from the complete decoupling of "inner" and spacetime integrals and their treatments.

\subsection{Inclusion of Standard Model "Matter" Fields}
As discussed in \cite{chw2} Isometrodynamics interacts with all fundamental fields appearing in a QFT such as the Standard Model (SM) of elementary particle physics through minimal coupling. For clarity we call all these fundamental other scalar, spinor and (gauge) vector fields "matter" fields in the sequel. For a potential physical interpretation of Isometrodynamics it is hence crucial to extend the analysis of the asymptotic scaling behaviour above to include the impact of these other fields on the renormalized coupling and the $\be$-function.

To be specific let us do this analysis for the SM fields which we minimally couple to Isometrodynamics by (1) allowing all SM fields to live on ${\bf M\/}^{\sl 4}\times {\bf R\/}^{D}$ - adding the necessary additional "inner" degrees of freedom -  and by (2) replacing ordinary derivatives through covariant ones $\pa_\mu\ar D_\mu = \pa_\mu + A_\mu\,^K\cdot \nabla_K $ in all "matter" Lagrangians as usual. 

In Appendix B we have derived the additional divergent contributions ${\it\Delta}\Ga_{\La, \sl{1}-loop}^{div} \left[A \right]$ to the one-loop effective action contributing to the renormalization of Isometrodynamcs.

To apply this to the SM let us recall its field content. The SM is built on gauging $SU(3)\times SU(2)\times U(1)$ which leaves us with $8$ strongly, $3$ weakly and $1$ electromagnetically interacting gauge fields - $12$ in total. These fields interact with $3$ families of leptons and quarks, two of which are structural replications of the first family consisting of the $15$ chiral Dirac fields for $\nu_e, e_L, e_R, u^\al_L, u^\al_R, d^\al_L, d^\al_R$, where $\al = 1,2,3$ indicates the strongly interacting color degrees of freedom. Finally there is a Higgs dublett adding two scalar degrees of freedom.

In total we have
\beqq \label{93}
& & \Ga_{\La, \sl{1}-loop}^{div} \left[A \right] \ar 
\Ga_{\La, \sl{1}-loop}^{div} \left[A \right] +
\,12\, {\it\Delta}_G \! \Ga_{\La, \sl{1}-loop}^{div} \left[A \right]
\nonumber \\
& & \quad\quad + \,\,45\, {\it\Delta}_D \! \Ga_{\La, \sl{1}-loop}^{div} \left[A \right] + \,2\, {\it\Delta}_S \! \Ga_{\La, \sl{1}-loop}^{div} \left[A \right] \\
& & = \,\frac{\Om_4}{\ep} \,\frac{\Om_D}{D(D+2)} \,\frac{1}{12} \,
\Big(11D + 24 - 90 - 2 \Big)
\,\La^2 \,\int \La^D \,F_{\mu\nu}\,^M \cdot F^{\mu\nu}\,_M, \nonumber
\eeqq
where $24$ is the contribution of the SM gauge fields, $90$ of the leptons and quarks and $2$ of the Higgs respectively. This translates into the renormalized coupling
\beq \label{94}
g_R = g\left(1 + \frac{g^2}{4 \pi^2} \,\frac{\Om_D}{D(D+2)} \,
\,\frac{1}{12} \, \Big(11(D - 6) - 2 \Big)\,
\frac{1}{\ep} + O(g^4)\right)
\eeq
and the $\be$-function
\beq \label{95}
\be(g) = - \frac{g^3}{4 \pi^2} \,\frac{\Om_D}{D(D+2)} 
\,\frac{1}{12} \, \Big(11(D - 6) - 2 \Big)
\eeq
of Isometrodynamics coupled to the Standard Model fields.

The combined theory is asymptotically free for $11(D - 6) - 2 > 0$ or for $D \geq 7$. If the flip-side of asymptotic freedom is confinement the "inner" space degrees of freedom and the gauge and "matter" fields associated with them are not expected to be directly observable - much as the gluons and quarks in QCD. 

Only for $2 \leq D \leq 6$ do we expect the "inner" space degrees of freedom and the gauge and "matter" fields associated with them to be observable and asymptotic states to exist. In this case it makes sense to evaluate the classical limit of Isometrodynamics.

Note that in the absence of the Higgs field the combined theory at one loop for $D=6$ is not renormalized at all.

\section{BRST Symmetry and BRST Quantization}
In this section we introduce the nilpotent BRST transformations for Isome-trodynamics and establish the BRST invariance of the gauge-fixed action. We define the physical states as equivalence classes of states in the kernel of the nilpotent BRST operator $Q$ modulo the image of $Q$. Finally we discuss the generalized BRST quantization of Isometrodynamics.

Let us start with the modified action $S_{MOD}$ from Eqn.(\ref{31}) which may be written as
\beq \label{96}
S_{MOD} = S_{ID} - \frac{\La^2}{2\xi} \, \int\,f_R \cdot f^R
+ \La^2 \, \int \, \om^*_R \cdot {\it\Delta}^R,
\eeq
where we have introduced the quantity
\beq \label{97}
{\it\Delta}^R \equiv {\cal F}^R\,_S \, \om^S.
\eeq 
Next we reexpress
\beqq \label{98}
B[f] &=& \exp\left\{ - i\, \frac{\La^2}{2\xi} \, \int\,f_R \cdot f^R \right\} \\
&\propto& \int\Pi_{\!\!\!\!\!\!_{_{_{x,X,R}}}}\!\!\!\!dh^R \;
\de (\nabla_R h^R)
\cdot \exp\left\{ i\, \frac{\La^2\xi}{2} \, \int\,h_R \cdot h^R
+ i \,\La^2 \, \int\,h_R \cdot f^R \right\}
\nonumber \eeqq
as a Gaussian integral and introduce the corresponding new modified action
\beq \label{99}
S_{NEW} = S_{ID} + \La^2 \, \int \, \om^*_R \cdot {\it\Delta}^R + \La^2 \, \int\,h_R \cdot f^R + \frac{\La^2\xi}{2} \, \int\,h_R \cdot h^R.
\eeq
Green functions are now given as path integrals over the fields $A$, $\om^*$, $\om$, $h$, $\psi$ with weight $\exp\, i \, \{S_{NEW}+S_M\}$.

By construction the gauge-fixed modified action $S_{NEW}$ is not invariant under gauge transformations. However, it is invariant under BRST transformations parametrized by an infinitesimal constant $\theta$ anticommuting with ghost and fermionic fields. The BRST variations are given by
\beqq \label{100}
\de_\theta A_\mu\,^M &=& \theta\left(\pa_\mu \om^M
+ A_\mu\,^K \nabla_K \om^M -\om^K \nabla_K A_\mu\,^M \right) \nonumber \\
\de_\theta \om^*_R &=& - \theta\, h_R \nonumber \\
\de_\theta \om^S &=& - \theta\, \om^K \nabla_K \om^S \\
\de_\theta h_R &=& 0 \nonumber \\
\de_\theta \psi &=& - \theta\, \om^K \nabla_K \psi.
\nonumber \eeqq
The transformations Eqns.(\ref{100}) are nilpotent, i.e. if ${\cal F}$ is any functional of $A, \om^*, \om, h, \psi$ and we define $s{\cal F}$ by
\beq \label{101}
\de_\theta {\cal F} \equiv \theta s{\cal F}
\eeq
then  
\beq \label{102}
\de_\theta s{\cal F} = 0 \quad\mbox{or}\quad s(s{\cal F}) = 0.
\eeq 
The proof for the fields above is straightforward, but somewhat tedious. Here we just give a scetch ot the verification of $s(s A_\mu\,^M)=0$
\beqq \label{103}
\de_\theta sA_\mu\,^M &=& \theta\, \Bigg\{
\pa_\mu \left( - \om^K \nabla_K \om^M \right) \nonumber \\
&+& \left(\pa_\mu \om^K
+ A_\mu\,^L \nabla_L \om^K -\om^L \nabla_L A_\mu\,^K \right) \nabla_K \om^M \nonumber \\
&-& A_\mu\,^K \nabla_K \left( \om^L \nabla_L \om^M \right)
+ \left( \om^L \nabla_L \om^K \right) \nabla_K A_\mu\,^M \\
&+& \om^K \nabla_K \left( \pa_\mu \om^M
+ A_\mu\,^L \nabla_L \om^M - \om^L \nabla_L A_\mu\,^M \right)\Bigg\} \nonumber \\
&=& 0 \nonumber \eeqq
using the chain-rule and the anticommutativity of $\theta$ with $\om$. As a result we have
\beqq \label{104}
s(s A_\mu\,^M) &=& 0,\quad s(s \om^*_R) = 0, \quad s(s \om^S) = 0 \nonumber \\
s(s h_R) &=& 0, \quad s(s\psi) = 0.
\eeqq
The extension to products of polynomials in these fields follows then easily.

To verify the BRST invariance of $S_{NEW}$ we note that the BRST transformation acts on functionals of matter and gauge fields alone as a gauge transformation with gauge parameter ${\cal E}_M = \theta\, \om_M$. Hence
\beq \label{105}
\de_\theta S_{ID} = 0.
\eeq
Next with the use of Eqn.(\ref{20}) we determine the BRST transform of $f^R$
\beq \label{106}
\de_\theta f^R = \frac{\de f^R}{\de\, {\cal E}_M} _{\mid_{_{{\cal E}=0}}} \!\!\!\! \theta \, \om_M= \theta \, {\it\Delta}^R
\eeq
which yields
\beq \label{107}
\om^*_R \cdot {\it\Delta}^R + h_R \cdot f^R + \frac{\xi}{2} \, h_R \cdot h^R = -s\left(\om^*_R \cdot f^R + \frac{\xi}{2} \, \om^*_R \cdot h^R \right).
\eeq
Hence we can rewrite
\beq \label{108}
S_{NEW} = S_{ID} + s{\it\Psi},
\eeq
where
\beq \label{109}
{\it\Psi} \equiv -\La^2 \, \int \left(\om^*_R \cdot f^R + \frac{\xi}{2} \, \om^*_R \cdot h^R \right).
\eeq
It finally follows from the nilpotency of the BRST transformation
\beq \label{110}
\de_\theta S_{NEW} = 0.
\eeq

As for Yang-Mills theories Eqn.(\ref{108}) shows that the physical content of Isometrodynamics is contained in the kernel of the BRST transformation modulo terms in its image.

Equivalent to this is the requirement that matrix elements between physical states $\mid\! \alpha\rangle,\dots$ are independent of the choice of the gauge-fixing functional ${\it\Psi}$. This implies the existence of a nilpotent BRST operator $Q$ with $Q^2=0$. Physical states are then in the kernel of $Q$
\beq \label{111}
Q \!\mid\! \alpha\rangle = 0,\quad \langle \beta \!\mid\! Q = 0.
\eeq
Independent physical states are defined as the equivalence classes of states in the kernel of $Q$ modulo the image of $Q$.

Finally let us note that as for Yang-Mills theories \cite{stw2} we can generalize the Faddeev-Popov-de Witt quantization procedure. In the general case one starts with an action given as the most general local functional of $A$, $\om^*$, $\om$, $h$, $\psi$ with ghost number zero which is invariant under the BRST transformations Eqns.(\ref{100}) and any other global symmetry of the theory as well as with dimension less or equal to four so as to assure renormalizability. Such actions are of the general form \cite{stw2}
\beq \label{112}
S_{NEW} [A,\om^*,\om,h,\psi]  = S_{ID} [\phi] + s{\it\Psi} [A,\om^*,\om,h,\psi]
\eeq
with $s{\it\Psi}$ being a general functional respecting the restrictions above.

$S$-matrix elements of physical states annihilated by the appropriate BRST operator of the theory are then independent of ${\it\Psi}$. In addition, in the Euclidean plus axial gauge the ghosts decouple in QID, hence they decouple for any choice of ${\it\Psi}$ and physical Isometrodynamics is ghost-free.

\section{Renormalizability to All Orders}
In this section we scetch a proof of the renormalizability of Isometrodynamics to all orders.

A general proof of the renormalizability of Isometrodynamics, i.e. the existence of a finite, well-defined perturbative effective action, has to comprise the analysis of the divergence structure and the renormalizability of spacetime integrals as for Yang-Mills theories and in addition the verification that "inner" space integrals can be properly defined respecting the scale invariance of the classical theory.

Turning to the first point we note that we should be able to employ the full machinery developed for the inductive renormalizability proof for Yang-Mills gauge theories as the general structure of Quantum Isometrodynamics formally is close to that of quantum Yang-Mills theories. Hence we should be able to repeat all the steps in the renormalizability proof e.g. given in the Chapters 15 to 17 in \cite{stw2} or in \cite{jez}. The only change arises from the slightly different form of the BRST transformations for Isometrodynamics as compared to Yang-Mills gauge theories requiring the adaptation of the analysis given in Section 17.2 of \cite{stw2}.

Turning to the second point our approach at the one-loop level has been to
(1) define the "inner" one-loop integrals using $\La$ as a cut-off
\beq \label{113}
\int \! \frac{d^{D}P}{(2{\pi})^D} \times \mbox{integrand} \ar
\int_{\mid P\mid \leq \La} \! \frac{d^{D}P}{(2{\pi})^D} \times \mbox{integrand}
\eeq
and (2) on the basis of this definition to demonstrate the validity of the scaling law
\beq \label{114}
\Ga^{(1-loop)}_{\rho\La} (X,A_\nu\,^M(X),\dots) =
\Ga^{(1-loop)}_{\La} (\rho X,\rho A_\nu\,^M(X),\dots)
\eeq
ensuring the uniqueness of the theory up to "inner" rescalings.

The same strategy should work for any number of loops. Again (1) we {\it define} "inner" $n$-loop integrals by
\beqq \label{115}
& & \int \! \frac{d^{D}P_1}{(2{\pi})^D} \cdot\dots\cdot \! \frac{d^{D}P_n}{(2{\pi})^D} \times \mbox{integrand} \\
& & \quad \ar \,\int_{\mid P_1\mid \leq \La} \! \frac{d^{D}P_1}{(2{\pi})^D} \cdot\dots\cdot \int_{\mid P_n\mid \leq \La} \! \frac{d^{D}P_n}{(2{\pi})^D} 
\times \mbox{integrand}
\nonumber \eeqq
arising in the calculation of the effective action and (2) on the basis of this definition we should be able to demonstrate the validity of the scaling law
\beq \label{116}
\Ga^{(n-loop)}_{\rho\La} (X,A_\nu\,^M(X),\dots) =
\Ga^{(n-loop)}_{\La} (\rho X,\rho A_\nu\,^M(X),\dots)
\eeq
noting that the "inner" scale invariance is a {\it linearly realized} symmetry of Isometrodynamics and hence a symmetry of the quantum effective action \cite{stw2}. This ensures the uniqueness of the theory up to "inner" rescalings at $n$ loops.

The locality of the theory in "inner" space for any number of loops follows from the non-propagation of "inner" degrees of freedom which can be most easily read off the propagators in Eqns.(\ref{41}).

This completes the scetch of a general proof of the renormalizability and the essential uniqueness of Quantum Isometrodynamics.

\section{Conclusions}
In this paper we have developed Isometrodynamics at the quantum level.

Important aspects of the quantization have been dealt with in very close analogy to the quantization of Yang-Mills gauge theories - in particular all the aspects related to (1) dealing with the pure gauge degrees of freedom, to (2) developing a perturbation theory and to spacetime-related divergencies of Feynman integrals and to (3) the asymptotic behaviour of the theory. The generalization of all these aspects to the infinitely many "inner" degrees of freedom of QID have posed no fundamentally new problems.

However, there were new challenges to be adressed - all related to the gauge group ${\overline{DIFF}}\,{\bf R}^D$. First, the gauge field and ghost variables $A_\mu\,^M$ and $\om^*_R$, $\om^S$ which naturally emerge from the gauging program had to be subject to restrictions which ensure that they live in the appropriate gauge algebra ${\overline{\bf diff}}\,{\bf R}^D$. These restrictions are not constraints in the usual sense and had to be implemented in the theory in a consistent way, i.e. through constraints in the gauge field and ghost functional measures for the path integrals for the theory's Green functions. Second, the gauge group is not compact and to avoid divergencies related to the infinite group volume we had to properly deal with the "sums over inner degrees of freedom". They become integrals over "inner" coordinates in Isometrodynamics and had to be defined through a regularization procedure respecting the "inner" scale invariance of the theory at the quantum level - making QID unique up to "inner" rescalings. Third, the "inner" degrees of freedom, in particular the "inner" global translation invariance of Isometrodynamics and the related conserved "inner" momenta have brought along new quantum numbers which will eventually require interpretation. 

As an overall result Isometrodynamics viewed as a generalization of non-Abelian gauge theories of compact Lie groups seems to stand on a solid basis as a new type of renormalizable gauge theory.

But there is an obvious question here. Can QID be used to describe fundamental interactions in Nature at both the classical and quantum level with gravity being a central candidate? Or is QID just a mathematical generalization of a framework which is limited to successfully describe the strong, weak and electromagnetic interactions in Nature?

To shed some light on this question let us point out that by its very definition Isometrodynamics fulfills important requirements towards any theory of gravity such as universality, i.e. the universal coupling of gravity to all fundamental particles and fields or such as the existence of a sensible limit for the case of gravity being "turned off", i.e. the laws of physics reducing to the Standard Model of elementary particle physics or a generalization thereof.

We will separately discuss the potential of Isometrodynamics to be a theory of gravity.

\appendix

\section{Dimensional Regularization of Divergent One-Loop Integrals}
In this Appendix we calculate the divergent contributions to the one-loop functional determinant Eqn.(\ref{78}) in four spacetime dimensions. $ _{\mid_{div}}$ indicates that we will retain only divergent terms and discard all finite contributions in a calculation.

\subsection{The contribution $\Ga^{(1)\, div}_\La$ linear in the fields}

\beq \label{117}
\Ga^{(1)\, div }_\La = \int\! d^4 x_1\int\! \frac{d^4 p_1}{(2{\pi})^4} {\Tr_{\!\!\!\!\!\!\!_{_{_{X}}}}}_\La \: \Bigg\{
\frac{1}{p_1^2} \,
\Big( i\,{\cal B}_{\rho_1} (x_1) \,p_1^{\rho_1} + {\cal C} (x_1) \Big)
\Bigg\}_{\mid_{div}}
\eeq
which identically vanishes using dimensional regularization,
\beq \label{118}
\frac{1}{(2{\pi})^4} \int\! d^4 p \, \frac{1}{p^2}
\ar \frac{i}{(2{\pi})^d} \int\! d^d p \, \frac{1}{p^2} = 0
\eeq
For the evaluation we have Wick-rotated the integral $p^{\sl 0}\ar i p^{\sl 4}$ and ana- lytically continued it to $d=4-\ep$ dimensions as we will do for all the integrals below.

\subsection{The contribution $\Ga^{(2)\, div}_\La$ quadratic in the fields}

\beqq \label{119}
\Ga^{(2)\, div }_\La &=& \int\! d^4 x_1\int\! d^4 x_2
\int\! \frac{d^4 p_1}{(2{\pi})^4}\!
\int\! \frac{d^4 k_2}{(2{\pi})^4} \nonumber \\
& & \!\!\!\!\!{\Tr_{\!\!\!\!\!\!\!_{_{_{X}}}}}_\La \: \Bigg\{
\frac{1}{p_1^2} \,
\Big( i\,{\cal B}_{\rho_1} (x_1) \,p_1^{\rho_1} + {\cal C} (x_1) \Big)
\nonumber \\
& & \cdot\quad \frac{1}{(p_1 + k_2)^2} \,
\Big( i\,{\cal B}_{\rho_2} (x_2) \,\left(p_1^{\rho_2} + k_2^{\rho_2} \Big) + {\cal C} (x_2) \right) \Bigg\}_{\mid_{div}} \nonumber \\
& & \!\!\!\!\!\cdot\exp \left( -i x_1 k_2 + i x_2 k_2 \right) \nonumber \\
&=& \int\! d^4 x_1\int\! d^4 x_2
\int\! \frac{d^4 p_1}{(2{\pi})^4}\!
\int\! \frac{d^4 k_2}{(2{\pi})^4} \\
& & \!\!\!\!\!{\Tr_{\!\!\!\!\!\!\!_{_{_{X}}}}}_\La \: \Bigg\{
\frac{p_1^\mu \, p_1^\nu}{p_1^2 (p_1 + k_2)^2} \,
\Big( i\,{\cal B}_\mu (x_1) \, i\,{\cal B}_\nu (x_2) \Big)
\nonumber \\
& & + \quad \frac{p_1^\mu}{p_1^2 (p_1 + k_2)^2} \,
\Big( i\,{\cal B}_\mu (x_1) \, i\,{\cal B}_\rho (x_2) \, k_2^\rho 
\nonumber \\
& & \quad\quad + \quad i\,{\cal B}_\mu (x_1) \,{\cal C} (x_2)
+ {\cal C} (x_1)\, i\,{\cal B}_\mu (x_2) \Big)
\nonumber \\
& & + \quad \frac{1}{p_1^2 (p_1 + k_2)^2} \,
\Big( {\cal C} (x_1) \, i\,{\cal B}_\rho (x_2) \, k_2^\rho
+ {\cal C} (x_1) \,{\cal C} (x_2) \Big) 
\Bigg\}_{\mid_{div}} \nonumber \\
& & \!\!\!\!\!\cdot\exp \left( -i x_1 k_2 + i x_2 k_2 \right). \nonumber 
\eeqq
We evaluate
\beqq \label{120}
\frac{1}{(2{\pi})^4} \int\! d^4 p \, \frac{1}{ p^2 (p + k_2)^2} _{\mid_{div}}
&\ar& \, i\,\frac{\Om_4}{\ep} \\ \label{120}
\frac{1}{(2{\pi})^4} \int\! d^4 p \, \frac{p^\mu}{p^2 (p + k_2)^2} _{\mid_{div}} &\ar& \, - i \,\frac{k_2^\mu}{2} \,\frac{\Om_4}{\ep} \\ \label{121}
\frac{1}{(2{\pi})^4} \int\! d^4 p \, \frac{p^\mu \, p^\nu}{p^2 (p + k_2)^2} _{\mid_{div}} &\ar& \left(\frac{i}{3} \, k_2^\mu \, k_2^\nu - \frac{i}{12} \, \eta^{\mu\nu} k_2^2 \right) \,\frac{\Om_4}{\ep} 
\eeqq
and obtain
\beqq \label{123}
\Ga^{(2)\, div }_\La &=& \, i\,\frac{\Om_4}{\ep} \, \intx_1 \,
{\Tr_{\!\!\!\!\!\!\!_{_{_{X}}}}}_\La \: \Bigg\{
\frac{1}{6}\, \pa^\mu\,{\cal B}_\mu (x_1) \cdot \pa^\nu\,{\cal B}_\nu (x_1) \nonumber \\
& & + \quad \frac{1}{12}\, \pa^\nu\,{\cal B}_\mu (x_1) \cdot 
\pa_\nu\,{\cal B}^\mu (x_1) \\
& & - \quad \pa^\mu\,{\cal B}_\mu (x_1) \cdot {\cal C} (x_1)
- {\cal C}^2 (x_1) \Bigg\} \nonumber
\eeqq 
after rewriting $k_2^\mu \exp(i x_2 k_2) = -i\,{\pa\rvec}^\mu_2 \exp(i x_2 k_2)$, partially integrating ${\pa\rvec}^\mu_2$, integrating out $k_2$, $x_2$ and using the cyclicality of the trace which is easily shown to hold also true in the case of ${\cal B}_\mu$, ${\cal C}$ being matrix-valued differential operators acting on "inner" space coordinates.

\subsection{The contribution $\Ga^{(3)\, div}_\La$ cubic in the fields}

\beqq \label{124}
\Ga^{(3)\, div }_\La &=& \int\! d^4 x_1\int\! d^4 x_2\int\! d^4 x_3
\int\! \frac{d^4 p_1}{(2{\pi})^4}\!
\int\! \frac{d^4 k_2}{(2{\pi})^4}\!
\int\! \frac{d^4 k_3}{(2{\pi})^4} \nonumber \\
& & \!\!\!\!\!{\Tr_{\!\!\!\!\!\!\!_{_{_{X}}}}}_\La \: \Bigg\{
\frac{1}{p_1^2} \,
\Big( i\,{\cal B}_{\rho_1} (x_1) \,p_1^{\rho_1} + {\cal C} (x_1) \Big)
\nonumber \\
& & \cdot\quad \frac{1}{(p_1 + k_2)^2} \,
\Big( i\,{\cal B}_{\rho_2} (x_2) \,\left(p_1^{\rho_2} + k_2^{\rho_2} \Big) + {\cal C} (x_2) \right) \nonumber \\
& & \cdot\quad \frac{1}{(p_1 + k_2 + k_3)^2} \,
\Big( i\,{\cal B}_{\rho_3} (x_3) \,\left(p_1^{\rho_3} + k_2^{\rho_3} + k_3^{\rho_3} \Big) + {\cal C} (x_3) \right) \Bigg\}_{\mid_{div}} \nonumber \\
& & \!\!\!\!\!\cdot\exp \left( -i x_1 (k_2 + k_3) + i x_2 k_2 + i x_3 k_3 \right) \nonumber \\
&=& \int\! d^4 x_1\int\! d^4 x_2\int\! d^4 x_3
\int\! \frac{d^4 p_1}{(2{\pi})^4}\!
\int\! \frac{d^4 k_2}{(2{\pi})^4}\!
\int\! \frac{d^4 k_3}{(2{\pi})^4} \\
& & \!\!\!\!\!{\Tr_{\!\!\!\!\!\!\!_{_{_{X}}}}}_\La \: \Bigg\{
\frac{p_1^\mu \, p_1^\nu \, p_1^\rho}{p_1^2 (p_1 + k_2)^2 (p_1 + k_2 + k_3)^2} \,
\Big( i\,{\cal B}_\mu (x_1) \, i\,{\cal B}_\nu (x_2) \, i\,{\cal B}_\rho (x_3)  \Big) \nonumber \\
& & + \quad \frac{p_1^\mu \, p_1^\nu}{p_1^2 (p_1 + k_2)^2 (p_1 + k_2 + k_3)^2} \, \Big( i\,{\cal B}_\mu (x_1) \, i\,{\cal B}_\nu (x_2) \,{\cal C} (x_3) 
\nonumber \\
& & \quad\quad + \quad
i\,{\cal B}_\mu (x_1) \,{\cal C} (x_2) \, i\,{\cal B}_\nu (x_3) 
+ i\,{\cal C} (x_1) \,{\cal B}_\mu (x_2) \, i\,{\cal B}_\nu (x_3) 
\nonumber \\
& & \quad\quad + \quad
i\,{\cal B}_\mu (x_1) \, i\,{\cal B}_\nu (x_2) 
i\,{\cal B}_\si (x_3) (k_2^\si + k_3^\si)
\nonumber \\
& & \quad\quad + \quad
i\,{\cal B}_\mu (x_1) i\,{\cal B}_\si (x_2) k_2^\si \, i\,{\cal B}_\nu (x_3)
\Big) \Bigg\}_{\mid_{div}} \nonumber \\
& & \!\!\!\!\!\cdot\exp \left( -i x_1 (k_2 + k_3) + i x_2 k_2 + i x_3 k_3 \right). \nonumber 
\eeqq
After evaluating
\beqq \label{125}
& & \frac{1}{(2{\pi})^4} \int\! d^4 p \, \frac{p^\mu \, p^\nu}
{p^2 (p + k_2)^2 (p + k_2 + k_3)^2} _{\mid_{div}} \ar 
\, \frac{i}{4} \, \eta^{\mu\nu} \,\frac{\Om_4}{\ep} \\ \label{125}
& & \frac{1}{(2{\pi})^4} \int\! d^4 p \, \frac{p^\mu \, p^\nu \, p^\rho}
{p^2 (p + k_2)^2 (p + k_2 + k_3)^2} _{\mid_{div}} \\ \label{126}
& & \quad\ar - \frac{i}{12} \Big( \eta^{\mu\nu} (2 k_2^\rho + k_3^\rho) + \eta^{\nu\rho} (2 k_2^\mu + k_3^\mu) + \eta^{\rho\mu} (2 k_2^\nu + k_3^\nu) \Big) \,\frac{\Om_4}{\ep} \nonumber
\eeqq
we obtain
\beqq \label{127}
\Ga^{(3)\, div }_\La &=& \, i\,\frac{\Om_4}{\ep} \, \intx_1 \,
{\Tr_{\!\!\!\!\!\!\!_{_{_{X}}}}}_\La \: \Bigg\{
\frac{1}{4}\, \pa^\mu\,{\cal B}_\mu (x_1) \cdot {\cal B}^\nu (x_1) \cdot {\cal B}_\nu (x_1) \\
& & - \quad \frac{1}{4}\, {\cal B}_\mu (x_1) \cdot \pa^\mu {\cal B}^\nu (x_1) \cdot {\cal B}_\nu (x_1) 
- \frac{3}{4}\, {\cal C} (x_1) \cdot {\cal B}^\nu (x_1) \cdot {\cal B}_\nu (x_1) \Bigg\}. \nonumber
\eeqq 
Here we have rewritten $k_j^\mu \exp(i x_j k_j) = -i\,{\pa\rvec}^\mu_j \exp(i x_j k_j)$, partially integrated ${\pa\rvec}^\mu_j$, integrated out $k_j$, $x_j$ for both $j=2,3$ and used the cyclicality of the trace.

\subsection{The contribution $\Ga^{(4)\, div}_\La$ quartic in the fields}

\beqq \label{128}
\Ga^{(4)\, div }_\La &=& \int\! d^4 x_1\int\! d^4 x_2
\int\! d^4 x_3\int\! d^4 x_4
\int\! \frac{d^4 p_1}{(2{\pi})^4}\!
\int\! \frac{d^4 k_2}{(2{\pi})^4}\!
\int\! \frac{d^4 k_3}{(2{\pi})^4}\!
\int\! \frac{d^4 k_4}{(2{\pi})^4} \nonumber \\
& & \!\!\!\!\!{\Tr_{\!\!\!\!\!\!\!_{_{_{X}}}}}_\La \: \Bigg\{
\frac{1}{p_1^2} \,
\Big( i\,{\cal B}_{\rho_1} (x_1) \,p_1^{\rho_1} + {\cal C} (x_1) \Big)
\nonumber \\
& & \cdot\quad \frac{1}{(p_1 + k_2)^2} \,
\Big( i\,{\cal B}_{\rho_2} (x_2) \,\left(p_1^{\rho_2} + k_2^{\rho_2} \Big) + {\cal C} (x_2) \right) \nonumber \\
& & \cdot\quad \frac{1}{(p_1 + k_2 + k_3)^2} \,
\Big( i\,{\cal B}_{\rho_3} (x_3) \,\left(p_1^{\rho_3} + k_2^{\rho_3} + k_3^{\rho_3} \Big) + {\cal C} (x_3) \right) \nonumber \\
& & \cdot\quad \frac{1}{(p_1 + k_2 + k_3 + k_4)^2} \,
\Big( i\,{\cal B}_{\rho_4} (x_4) \\
& & \quad\quad\quad\quad\quad \cdot \left(p_1^{\rho_4} + k_2^{\rho_4} + k_3^{\rho_4} + k_4^{\rho_4} \Big) + {\cal C} (x_4) \right) \Bigg\}_{\mid_{div}} \nonumber \\
& & \!\!\!\!\!\cdot\exp \left( -i x_1 (k_2 + k_3 + k_4) + i x_2 k_2 + i x_3 k_3 + i x_4 k_4 \right) \nonumber \\
&=& \int\! d^4 x_1\int\! d^4 x_2
\int\! d^4 x_3\int\! d^4 x_4
\int\! \frac{d^4 p_1}{(2{\pi})^4}\!
\int\! \frac{d^4 k_2}{(2{\pi})^4}\!
\int\! \frac{d^4 k_3}{(2{\pi})^4}\!
\int\! \frac{d^4 k_4}{(2{\pi})^4} \nonumber \\
& & \!\!\!\!\!{\Tr_{\!\!\!\!\!\!\!_{_{_{X}}}}}_\La \: \Bigg\{
\frac{p_1^\mu \, p_1^\nu \, p_1^\rho \, p_1^\si }{p_1^2 (p_1 + k_2)^2 (p_1 + k_2 + k_3)^2 (p_1 + k_2 + k_3 + k_4)^2} \nonumber \\
& & \cdot\quad \Big( i\,{\cal B}_\mu (x_1) \, i\,{\cal B}_\nu (x_2) \, i\,{\cal B}_\rho (x_3)  \, i\,{\cal B}_\si (x_4) \Big) \Bigg\}_{\mid_{div}}
\nonumber \\
& & \!\!\!\!\!\cdot\exp \left( -i x_1 (k_2 + k_3 + k_4) + i x_2 k_2 + i x_3 k_3 + i x_4 k_4 \right). \nonumber 
\eeqq
With the use of
\beqq \label{129}
& & \frac{1}{(2{\pi})^4} \int\! d^4 p \,
\frac{p^\mu \, p^\nu \, p^\rho \, p^\si}
{p^2 (p + k_2)^2 (p + k_2 + k_3)^2 (p + k_2 + k_3 + k_4)^2} _{\mid_{div}} \nonumber \\
& & \quad \ar \frac{i}{24} \Big( \eta^{\mu\nu} \eta^{\rho\si} + \eta^{\mu\rho} \eta^{\nu\si} + \eta^{\mu\si} \eta^{\nu\rho} \Big) \,\frac{\Om_4}{\ep} 
\eeqq
we obtain
\beqq \label{130}
\Ga^{(4)\, div }_\La &=& \, i\,\frac{\Om_4}{\ep} \, \intx_1 \,
{\Tr_{\!\!\!\!\!\!\!_{_{_{X}}}}}_\La \: \Bigg\{
\frac{1}{12}\, {\cal B}^\mu (x_1) \cdot {\cal B}_\mu (x_1) \cdot {\cal B}^\nu (x_1) \cdot {\cal B}_\nu (x_1) \nonumber \\
& & + \quad \frac{1}{24}\, {\cal B}^\mu (x_1) \cdot {\cal B}^\nu (x_1) \cdot {\cal B}_\mu (x_1) \cdot {\cal B}_\nu (x_1) \Bigg\}
\eeqq 
after integrating out $k_j$, $x_j$ for all $j=2,3,4$ and using the cyclicality of the trace.

\section{"Matter" Contributions to Divergent Part of One-Loop Effective Action of  Isometrodynamics}
In this Appendix we calculate the divergent vacuum contribution of a gauge vector field, a Dirac spinor and a complex scalar doublet to the one-loop effective action of Isometrodynamics.

\subsection{Gauge field contribution ${\it\Delta}_G \! \Ga_{\La, \sl{1}-loop}^{div} \left[A \right]$}
The vacuum amplitude of a Yang-Mills gauge field $B_\mu\,^\al$ with gauge algebra indices $\al, \be,..=1,..,\dim A$ minimally coupled to Isometrodynamics, where $\dim A$ is the dimension of the gauge algebra, is given by
\beqq \label{131}
{\cal Z}_G [A] &\equiv&
\int\Pi_{\!\!\!\!\!\!_{_{_{x,X,\al,\mu}}}}
\!\!\!\!\!\!\!\!dB_\mu\,^\al \;
\int\Pi_{\!\!\!\!\!\!_{_{_{x,X,\be}}}}\!\!d\om^*_\be \;
\int\Pi_{\!\!\!\!\!\!_{_{_{x,X,\ga}}}}\!\!d\om^\ga
\nonumber \\
& & \cdot \exp\,i\,\left\{S_{MOD} + \ep \mbox{-terms} \right\}, \eeqq
where $D_\mu^\al\,\!_\be [B] = \pa_\mu \de^\al\,\!_\be + C^\al\,_{\ga\be} B_\mu\,^\ga$ is the covariant derivative in the presence of a gauge field $B$, $C^\al\,_{\ga\be}$ the structure constants of the gauge algebra and $\om^*_\be$, $\om^\ga$ the ghost fields corresponding to the gauge-fixed action $S_{MOD}$
\beqq \label{132}
S_{MOD}&\equiv& S_{YM} + S_{GF} + S_{GH} \nonumber \\
S_{YM}&\equiv& -\frac{1}{4} \,\int \, {\overline G}_{\mu\nu}\,^\al \cdot {\overline G}^{\mu\nu}\,_\al \\
S_{GF}&\equiv& - \frac{1}{2\xi} \,\int \,
{\overline D}^\mu_{\al\be} [C] B_\mu\,^\be \cdot {\overline D}_\nu^\al\,\!_\ga [C] B^{\nu\ga} \nonumber \\
S_{GH}&\equiv& \int \, \om^*_\be \cdot {\cal F}^\be\,_\ga \left[B, C \right] \om^\ga. \nonumber
\eeqq
$C_\mu\,^\al$ appearing in the gauge-fixing and ghost terms is a background gauge field. Above we have minimally coupled the Yang-Mills field to Isometrodynamics replacing ordinary through covariant derivatives $\pa_\mu \ar D_\mu = \pa_\mu + A_\mu\,^K\cdot \nabla_K$ yielding
\beq \label{133}
D_\mu^\al\,\!_\be [B] \ar {\overline D}_\mu^\al\,\!_\be [B] = D_\mu \de^\al\,\!_\be + C^\al\,_{\ga\be} B_\mu\,^\ga,
\eeq
and introduced the field strength and the ghost fluctuation operator
\beqq \label{134}
{\overline G}_{\mu\nu}\,^\al &=& D_\mu B_\nu\,^\al - D_\nu B_\mu\,^\al +
C^\al\,_{\be\ga} \, B_\mu\,^\be \, B_\nu\,^\ga, \nonumber \\
{\cal F}^\be\,_\ga \left[B, C \right] &=&
{\overline D}_\mu^\be\,\!_\al [C] \, {\overline D}^{\mu \al}\,\!_\ga [B].
\eeqq
The bars over derivatives etc. indicate minimal coupling to Isometrodynamics.

Expanding $S_{MOD}$ around its stationary points $B_\mu\,^\al=C_\mu\,^\al =\om^*_\be=\om^\ga=0$ in the absence of source terms and performing the Gaussian integral gives
\beqq \label{135}
{\cal Z}_{G, \sl{1}-loop} [A]
&=& \int\Pi_{\!\!\!\!\!\!_{_{_{x,X,\al,\mu}}}}
\!\!\!\!\!\!\!\!d\de B_\mu\,^\al \;
\int\Pi_{\!\!\!\!\!\!_{_{_{x,X,\be}}}}\!\!d\de \om^*_\be \;
\int\Pi_{\!\!\!\!\!\!_{_{_{x,X,\ga}}}}\!\!d\de \om^\ga \nonumber \\
& & \cdot \exp \Bigg\{ - \frac{i}{2} \, \int \, \de B_\mu\,^\al \cdot
{\cal D}_{B,\xi}^{\mu\nu}\,_{\al\be}\, \de B_\nu\,^\be \\
& & \quad\quad - \: \, \int \, \de \om^*_\be \cdot {\cal D}_\om^\be\,_\ga\, \de \om^\ga \Bigg\} \nonumber \\
&=& \Det^{-1/2}\, {\cal D}_{B,\xi} \cdot \Det\, {\cal D}_\om, \nonumber
\eeqq
where
\beqq \label{136}
{\cal D}_{B,\xi}^{\mu\nu}\,_{\al\be} &\equiv& - \left(\eta ^{\mu\nu} \cdot D^\rho \, D_\rho \, + \left(1 - \frac{1}{\xi} \right)\,
D^\mu \, D^\nu - F^{\mu\nu} \right) \de_{\al\be} \nonumber \\
{\cal D}_\om^\be\,_\ga &\equiv& -\, D^\rho \, D_\rho \, \de^\be\,_\ga.
\eeqq

Taking everything together and evaluating the divergent contribution to the one-loop effective action with the use of Eqns.(\ref{136}), (\ref{85}) and (\ref{86}) for $\xi=1$ yields for each independent gauge field and associated ghost
\beq \label{137}
{\it\Delta}_G \! \Ga_{\La, \sl{1}-loop}^{div} \left[A \right]
= \,\frac{\Om_4}{\ep} \,\frac{\Om_D}{D(D+2)} \,\frac{1}{6} 
\,\La^2 \,\int \La^D \,F_{\mu\nu}\,^M \cdot F^{\mu\nu}\,_M,
\eeq
where we have discarded the factor $\dim A$ which accounts for the number of independent gauge fields. Note that such a term will reinforce asymptotic freedom. Note in addition that this formula also holds in the Abelian case where the ghost contribution in the presence of $A_\mu\,^M$ does not reduce to a field-independent determinant.

\subsection{Dirac spinor contribution ${\it\Delta}_D \! \Ga_{\La, \sl{1}-loop}^{div} \left[A \right]$}
The vacuum amplitude of a Dirac field minimally coupled to Isometrodynamics is given by
\beq \label{138}
{\cal Z}_D [A] \equiv
\int\Pi_{\!\!\!\!\!\!_{_{_{x,X}}}}\!\!d {\overline\psi} \;
\int\Pi_{\!\!\!\!\!\!_{_{_{x,X}}}}\!\!d \psi
\, \exp\,i\,\left\{S_D + \ep \mbox{-terms} \right\}, \eeq
where $\psi$ is a Dirac spinor and 
\beq \label{139}
S_D \equiv – \int \, {\overline\psi} \, ({\overline \Dsl} + m) \, \psi
\eeq
is the spinor action coupled to a Yang-Mills field through the covariant derivative $D_\mu [B] = \pa_\mu - i\, t_\al B_\mu\,^\al$. Here $t_\al$ is the generator of the gauge algebra in the fermion space.

Again we have minimally coupled the Dirac field to Isometrodynamics replacing ordinary through covariant derivatives $\pa_\mu \ar D_\mu = \pa_\mu + A_\mu\,^K\cdot \nabla_K$ yielding
\beq \label{140}
D_\mu [B] \ar {\overline D}_\mu [B] = D_\mu - i\, t_\al B_\mu\,^\al.
\eeq

Expanding $S_D$ around its stationary points ${\overline\psi}=\psi=B_\mu\,^\al=0$ in the absence of external sources and performing the Grassmann integral gives
\beqq \label{141}
{\cal Z}_{D, \sl{1}-loop} [A]
&=& 
\int\Pi_{\!\!\!\!\!\!_{_{_{x,X}}}}\!\!d\de {\overline\psi} \;
\int\Pi_{\!\!\!\!\!\!_{_{_{x,X}}}}\!\!d\de \psi 
\exp \left\{ - i \, \int \, \de {\overline\psi} \cdot
{\cal D}_\psi \, \de\psi \right\} \nonumber \\
&=& \Det^{1/2}\, {\cal D}^2_\psi, 
\eeqq
where
\beq \label{142}
{\cal D}^2_\psi = - \Dsl^2 =
- D^\rho \, D_\rho \, - \frac{1}{2} \, F^{\mu\nu} \ga_\mu \ga_\nu.
\eeq

Taking everything together and evaluating the divergent contribution to the one-loop effective action with the use of Eqns.(\ref{142}), (\ref{85}) and (\ref{86}) yields for each independent Dirac spinor
\beq \label{143}
{\it\Delta}_D \! \Ga_{\La, \sl{1}-loop}^{div} \left[A \right]
= - \,\frac{\Om_4}{\ep} \,\frac{\Om_D}{D(D+2)} \,\frac{1}{3}
\,\La^2 \,\int \La^D \,F_{\mu\nu}\,^M \cdot F^{\mu\nu}\,_M.
\eeq
Note that this will work against asymptotic freedom. Note in addition that a chiral Dirac fields contributes just half of the value above.

\subsection{Scalar doublet contribution ${\it\Delta}_S \! \Ga_{\La, \sl{1}-loop}^{div} \left[A \right]$}
The vacuum amplitude of a complex scalar doublet minimally coupled to Isometrodynamics is given by
\beq \label{144}
{\cal Z}_S [A] \equiv
\int\Pi_{\!\!\!\!\!\!_{_{_{x,X}}}}\!\!d \va^\dagger \;
\int\Pi_{\!\!\!\!\!\!_{_{_{x,X}}}}\!\!d \va
\, \exp\,i\,\left\{S_S + \ep \mbox{-terms} \right\}, \eeq
where $\va$ is a complex scalar doublet and 
\beq \label{145}
S_S \equiv – \int \, \left( ({\overline D}_\mu \va)^\dagger \cdot ({\overline D}_\mu \va) + V(\va^\dagger\cdot\va) \right)
\eeq
is the doublet coupled to the $SU(2)\times U(1)$ gauge bosons of the electro-weak interaction through the covariant derivative $D_\mu [B] = \pa_\mu - i\, {B\rvec}_\mu\cdot {t\rvec}_\va - i\, B_\mu\, y_\va$.

Again we have minimally coupled the scalar to Isometrodynamics replacing ordinary through covariant derivatives $\pa_\mu \ar D_\mu = \pa_\mu + A_\mu\,^K\cdot \nabla_K$ yielding
\beq \label{146}
D_\mu [B] \ar {\overline D}_\mu [B] = D_\mu - i\, {B\rvec}_\mu\cdot {t\rvec}_\va - i\, B_\mu\, y_\va.
\eeq

Expanding $S_S$ around one of its stationary points ${B\rvec}_\mu=B_\mu=0$ and  $\va^\dagger\cdot\va=const.$ and performing the Gaussian integral gives
\beqq \label{147}
{\cal Z}_{S, \sl{1}-loop} [A]
&=& 
\int\Pi_{\!\!\!\!\!\!_{_{_{x,X}}}}\!\!d\de \va^\dagger \;
\int\Pi_{\!\!\!\!\!\!_{_{_{x,X}}}}\!\!d\de \va 
\exp \left\{ - i \, \int \, \de \va^\dagger \cdot
{\cal D}_\va \, \de\va \right\} \nonumber \\
&=& \Det^{-1}\, {\cal D}_\va
\eeqq
where
\beq \label{148}
{\cal D}_\va = - D^\rho \, D_\rho \, + \frac{\de V(\va^\dagger\cdot\va)}
{\de\va^\dagger\, \de\va}.
\eeq

Taking everything together and evaluating the divergent contribution to the one-loop effective action with the use of Eqns.(\ref{148}), (\ref{85}) and (\ref{86}) yields for a complex scalar doublet
\beq \label{149}
{\it\Delta}_S \! \Ga_{\La, \sl{1}-loop}^{div} \left[A \right]
= - \,\frac{\Om_4}{\ep} \,\frac{\Om_D}{D(D+2)} \,\frac{1}{6}
\,\La^2 \,\int \La^D \,F_{\mu\nu}\,^M \cdot F^{\mu\nu}\,_M,
\eeq
which holds independent of whether the Higgs mechanism is in place or not and will work against asymptotic freedom. Note that a single complex scalar field contributes just half of the value above.

\section{Notations and Conventions}

Generally, small letters denote spacetime coordinates and parameters, capital letters coordinates and parameters in "inner" space.

Specifically, ({\bf M\/}$^{\sl 4}$,\,$\eta$) denotes $\sl{4}$-dimensional Minkowski spacetime with the Cartesian coordinates $x^\la,y^\mu,z^\nu,\dots\,$ and the spacetime metric $\eta=\mbox{diag}(-1,1,1,1)$. The small Greek indices $\la,\mu,\nu,\dots$ from the middle of the Greek alphabet run over $\sl{0,1,2,3}$. They are raised and lowered with $\eta$, i.e. $x_\mu=\eta_{\mu\nu}\, x^\nu$ etc. and transform covariantly w.r.t. the Lorentz group $SO(\sl{1,3})$. Partial differentiation w.r.t to $x^\mu$ is denoted by $\pa_\mu \equiv \frac{\pa\,\,\,}{\pa x^\mu}$. 
Small Latin indices $i,j,k,\dots$ generally run over the three spatial coordinates $\sl{1,2,3}$ \cite{stw1}.

({\bf R\/}$^{D}$,\,$g$) denotes a $D$-dimensional real vector space with coordinates $X^L, Y^M, Z^N,\dots\,$ and the flat metric $g_{MN}$ with signature $D$. The metric transforms as a contravariant tensor of Rank 2 w.r.t. ${\overline{DIFF}}\,{\bf R}^D$. Because Riem$(g) = 0$ we can always choose global Cartesian coordinates and the Euclidean metric $\de=\mbox{diag}(1,1,\dots,1)$. The capital Latin indices $L,M,N,\dots$ from the middle of the Latin alphabet run over $\sl{1,2},\dots,D$. They are raised and lowered with $g$, i.e. $X_M=g_{MN} X^N$ etc. and transform as vector indices w.r.t. ${\overline{DIFF}}\,{\bf R}^D$. Partial differentiation w.r.t to $X^M$ is denoted by $\nabla_M \equiv \frac{\pa\quad\,}{\pa X^M}$. 

The same lower and upper indices are summed unless indicated otherwise.

\section*{Acknowledgments}

This paper is dedicated to my daughters Alina and Sarah and to my wife Francoise who have helped me to keep the right perspectives on this work and who have patiently carried me through the sometimes emotional ups and downs on the stony way to Isometrodynamics.

\end{document}